\documentclass[11pt,letterpaper]{article}

\usepackage[letterpaper,margin=1in]{geometry}
\usepackage{amsmath,amssymb,amsthm,mathtools}
\usepackage{microtype}
\usepackage{xcolor}
\usepackage[colorlinks=true,linkcolor=blue!55!black,citecolor=green!45!black,urlcolor=blue!55!black]{hyperref}
\usepackage{enumitem}
\usepackage{authblk}

\newif\ifannotationcolors
\annotationcolorsfalse
\colorlet{keycolor}{orange!85!black}
\colorlet{explaincolor}{blue!70!black}

\DeclareRobustCommand{\key}[1]{{\ifannotationcolors\color{keycolor}\else\color{black}\fi#1}}
\DeclareRobustCommand{\explain}[1]{{\ifannotationcolors\color{explaincolor}\else\color{black}\fi#1}}

\ifannotationcolors\else
  \hypersetup{allcolors=black}
\fi
\theoremstyle{plain}
\newtheorem{theorem}{Theorem}[section]
\newtheorem{lemma}[theorem]{Lemma}
\newtheorem{proposition}[theorem]{Proposition}
\newtheorem{corollary}[theorem]{Corollary}
\newtheorem{conjecture}[theorem]{Conjecture}
\theoremstyle{definition}
\newtheorem{definition}[theorem]{Definition}

\newtheorem{remark}[theorem]{Remark}
\newtheorem{fact}[theorem]{Fact}

\DeclareMathOperator{\rank}{rank}
\DeclareMathOperator{\codim}{codim}
\DeclareMathOperator{\supp}{supp}
\DeclareMathOperator{\mult}{mult}

\DeclareMathOperator*{\Ex}{\mathbb{E}}
\newcommand{\F}{\mathbb{F}}
\newcommand{\Z}{\mathbb{Z}}
\newcommand{\R}{\mathbb{R}}
\newcommand{\Dxor}{\mathsf{D}_{\oplus}}
\newcommand{\Dcc}{\mathsf{D}^{\mathrm{cc}}}
\newcommand{\eps}{\varepsilon}
\newcommand{\half}{\tfrac{1}{2}}

\NewCommandCopy{\bareeqref}{\eqref}
\renewcommand{\eqref}[1]{\textup{Eq.~\bareeqref{#1}}}

\title{\bf Parity Decision Trees of Depth $O(4^{d}\log\|\widehat f\|_1)$\\ for Boolean Functions $f$ of $\F_2$-Degree $d$}

\author{Shengyu Zhang}
\affil{Tencent Quantum Lab}
\date{} 

\begin{document}
\pagenumbering{roman}
\begin{titlepage}
\maketitle
\thispagestyle{empty}

\begin{abstract}
\normalsize
Let $f\colon\F_2^n\to\{-1,+1\}$ have $\F_2$-degree $d\ge2$. We prove that its deterministic parity decision tree depth satisfies
\[
  \Dxor(f)= O(4^d \log_2\|\widehat f\|_1) = O(4^d \log_2 s(f)),
\]
where $\|\widehat f\|_1=\sum_\xi|\widehat f(\xi)|$ is the Fourier spectral norm and $s(f)$ is the Fourier sparsity. The bound is linear in the logarithm of the spectral norm for every fixed degree, with singly exponential dependence on $d$. This improves the previously best known bound of $O(2^{d^2/2} \log^{d-2} \|\widehat f\|_1)$ due to Tsang, Wong, Xie and Zhang (FOCS 2013). Our result establishes the log-rank conjecture for $\mathrm{XOR}$ functions of $\F_2$-degree $O(\log\log s(f))$, extending the previously accessible range $d=O(1)$. The proof combines field lifting, multiplicative derivatives, a multiplicity-based generic-rank comparison, a Schur-complement vanishing argument, and dimension-preserving descent to $\F_2$ to lower the degree on every coset of a common subspace.
\end{abstract}
\end{titlepage}
\pagenumbering{arabic}

\section{Introduction}

Communication complexity studies how much information parties must exchange to
compute a function when its inputs are distributed among them. Besides modelling
distributed computation, it is a fundamental tool for proving lower bounds in
models such as circuits, data structures and streaming algorithms~\cite{KushilevitzNisan97,RaoYehudayoff20}.

\subsection{The log-rank conjecture and \texorpdfstring{$\mathrm{XOR}$}{XOR} functions}

For a two-party Boolean function $F\colon\mathcal X\times\mathcal Y\to\{-1,1\}$ let
$\Dcc(F)$ denote its deterministic communication complexity and let $\rank(F)$ denote the
rank over $\R$ of the matrix $(F(x,y))_{x,y}$. Mehlhorn and Schmidt~\cite{MS82} observed
that $\log_2\rank(F)\le \Dcc(F)$. The log-rank conjecture of Lov\'asz and
Saks~\cite{LS88} asserts a converse up to a polynomial loss:
\begin{equation}\label{eq:logrank}
  \Dcc(F)\;\le\;\mathrm{polylog}\bigl(\rank(F)\bigr).
\end{equation}
After almost four decades the best general upper bounds remain exponentially far
from \eqref{eq:logrank}: Lovett~\cite{Lovett16} proved
$\Dcc(F)=O(\sqrt{\rank(F)}\,\log\rank(F))$, improved to $O(\sqrt{\rank(F)})$ by
Sudakov and Tomon~\cite{SudakovTomon}. In the other direction, near-quadratic
separations have been sharpened to the largest known separation
$\Dcc(F)\ge\Omega(\log^{2}\rank(F))$~\cite{BBGJK21,Pabbaraju26},
with the lossless bound due to Pabbaraju.

Because the general conjecture appears out of reach, attention has turned to the
special case of \emph{$\mathrm{XOR}$ functions}, studied in this context by Shi and
Zhang~\cite{ShiZhang}, by Montanaro and Osborne~\cite{MontanaroOsborne}, and later
by Lin and Zhang~\cite{LinZhang17} for functions with small alternating numbers. Given
$f\colon\F_2^n\to\{-1,1\}$ one sets
\[
  F(x,y)\;=\;f(x\oplus y),\qquad x,y\in\F_2^n .
\]
The communication matrix of $F$ is then diagonalised by the Fourier basis, and its rank
equals the \emph{Fourier sparsity}
\[
  s(f)\;=\;\bigl|\supp\widehat f\bigr|,\qquad
  \widehat f(\xi)=2^{-n}\sum_{x\in\F_2^n}f(x)(-1)^{\xi\cdot x}.
\]
Moreover $\Dcc(F)$ is tightly linked to a query model. A \emph{parity decision tree}
(PDT) is a decision tree each of whose internal nodes queries the value of a linear form
$a\cdot x$ over $\F_2$; write $\Dxor(f)$ for the minimum worst-case depth of a PDT
computing $f$ exactly. Simulating a parity query costs two bits of communication
(Alice sends $a\cdot x$, Bob replies with the branch bit), so
\begin{equation}\label{eq:pdt-to-cc}
  \Dcc(F)\;\le\;2\,\Dxor(f),
\end{equation}
and Hatami, Hosseini and Lovett~\cite{HHL18} proved the nontrivial converse
$\Dxor(f)=O\bigl(\Dcc(F)^{6}\bigr)$. The two quantities are therefore polynomially
equivalent, and the log-rank conjecture for $\mathrm{XOR}$ functions is equivalent to the
following purely Fourier-analytic statement.

\begin{conjecture}[log-rank conjecture for $\mathrm{XOR}$ functions~\cite{MontanaroOsborne,ZhangShi10,TWXZ13}]
\label{conj:xor-logrank}
For every $f\colon\F_2^n\to\{-1,1\}$ one has $\Dxor(f)\le\mathrm{polylog}\bigl(s(f)\bigr)$.
\end{conjecture}

The trivial lower bound is $\Dxor(f)\ge\half\log_2 s(f)$, because a depth-$D$ PDT
partitions $\F_2^n$ into at most $2^{D}$ affine subspaces and hence
$s(f)\le4^{D}$. Shpilka, Tal and Volk~\cite{STV17} proved
$\Dxor(f)=O(\|\widehat f\|_1^2\log_2 s)$, while Tsang, Wong, Xie and
Zhang~\cite{TWXZ13} obtained $O(\|\widehat f\|_1\log_2 s)$.
The best unconditional upper bound in terms of sparsity alone,
$\Dxor(f)=O(\sqrt{s})$, also follows from the work of Tsang, Wong, Xie and
Zhang~\cite{TWXZ13}; Mande and Sanyal~\cite{MandeSanyal} later gave an alternative proof.
On the lower-bound side, O'Donnell, Sun, Tan, Wright and
Zhao~\cite{OSTWZ14} exhibited functions with
$\Dxor(f)=\Omega\bigl((\log_2 s)^{\log_{3}6}\bigr)$, $\log_36\approx1.63$, so the
polynomial in Conjecture~\ref{conj:xor-logrank} cannot be taken linear. Several natural
routes to Conjecture~\ref{conj:xor-logrank} through additive structure of
$\supp\widehat f$ have recently been refuted by Hatami, Hosseini, Lovett and
Ostuni~\cite{HHLO24}.

\subsection{Degree as a parameter}

This paper studies the regime in which the \emph{$\F_2$-degree} of $f$ is small. Every
$f\colon\F_2^n\to\{-1,1\}$ can be written uniquely as $f=(-1)^{p}$ for a multilinear
polynomial $p\colon\F_2^n\to\F_2$; we write $d=\deg p$ for the degree of its algebraic
normal form. The parameter $d$ never exceeds $\log_2 s$ when $d\ge2$
(Lemma~\ref{lem:d-le-logs} below), so degree-sensitive bounds are statements about a
genuine sub-range of the problem rather than about an unrelated parameter.

The known degree-sensitive bound is due to Tsang, Wong, Xie and Zhang.

\begin{theorem}[\cite{TWXZ13}]\label{thm:twxz}
For every $f=(-1)^{p}$ with $\deg p=d\ge 3$ and $F(x,y)=f(x\oplus y)$,
\[
  \Dcc(F)\;=\;O\!\left(2^{d^{2}/2}\,\log^{\,d-2}\|\widehat f\|_1\right).
\]
In particular Conjecture~\ref{conj:xor-logrank} holds for $\mathrm{XOR}$ functions of
constant $\F_2$-degree.
\end{theorem}

Two features of Theorem~\ref{thm:twxz} limit its reach. The degree factor
$2^{d^{2}/2}$ is quasi-polynomial in $s$ once $d$ grows beyond $\sqrt{\log s}$, and --
more importantly -- the power $\log^{d-2}$ of the logarithm is super-polylogarithmic
for any $d=\omega(1)$. Hence Theorem~\ref{thm:twxz} proves
Conjecture~\ref{conj:xor-logrank} only for $d=O(1)$.

\subsection{Our result}

Our main theorem removes the growing power of the logarithm entirely, replaces the
degree factor by a single exponential
.

\begin{theorem}[Main theorem]\label{thm:main}
Let $p\colon\F_2^n\to\F_2$ have degree exactly $d\ge2$ and let $f=(-1)^{p}$. Then
\[
  \Dxor(f)\;\le\;1+\frac{2^{2d-1}-5}{3}\,\log_2\|\widehat f\|_1
  \;\le\;1+\frac{2^{2d-1}-5}{6}\,\log_2 s(f).
\]
In particular $\Dxor(f)=O\bigl(4^{d}\log_2 s(f)\bigr)$, and by \eqref{eq:pdt-to-cc} the
same bound holds for $\half\Dcc(F)$.
\end{theorem}

The restriction $d\ge2$ removes a genuine endpoint: a nonconstant affine phase has
$s=1$ but depth one, so no bound of the form $c\cdot\log_2 s$ can hold there. A uniform
statement covering all cases is $\Dxor(f)=O(1+4^{d}\log_2 s)$.

For $d=2$ the theorem reads $\Dxor(f)\le1+\half\log_2 s$, which is exactly optimal
(Proposition~\ref{prop:quadratic-exact}). For every fixed $d\ge4$ and $s\to\infty$ the
bound of Theorem~\ref{thm:main} is asymptotically stronger than
Theorem~\ref{thm:twxz}, since it is linear rather than of degree $d-2$ in
$\log\|\widehat f\|_1$. The qualitative consequence is the following extension of the range
in which Conjecture~\ref{conj:xor-logrank} is known.

\begin{corollary}\label{cor:range}
If $f=(-1)^{p}$ with $\deg p = O(\log\log_2 s(f))$, then
$\Dxor(f)\le\mathrm{polylog}\bigl(s(f)\bigr)$. Consequently the log-rank conjecture holds for all $\mathrm{XOR}$ functions of $\F_2$-degree $O(\log\log s)$.
\end{corollary}


The proof actually establishes a stronger statement which replaces the spectral norm by the \emph{threshold weight}
\[
  W(f)\;=\;\min\Bigl\{\|\widehat q\|_1 \;:\; q\colon\F_2^n\to\R,\ f(x)q(x)\ge1,\ \forall x\Bigr\}
  \;\le\;\|\widehat f\|_1 .
\]
Note that this $W(f)$ is always at most $\|\widehat f\|_1$ because one can simply choose $q$ to be $f$.
\begin{theorem}\label{thm:main-weight}
Under the hypotheses of Theorem~\ref{thm:main},
$\displaystyle \Dxor(f)\le1+\frac{2^{2d-1}-5}{3}\log_2 W(f)$.
\end{theorem}

\subsection{Techniques}\label{sec:techniques}

The proof follows the \emph{degree-reduction} paradigm introduced in~\cite{TWXZ13}: one
exhibits a subspace $U\le\F_2^n$ of small codimension on which the top-degree part of $p$
vanishes identically, queries a basis of its annihilator $U^{\perp}$, and recurses on
each of the resulting cosets, where the degree has dropped by at least one. Two facts make
the recursion affordable.

First, if the top-degree form vanishes on $U$ then the degree drops on \emph{every} coset
of $U$ at once, so all $2^{\codim U}$ branches make progress simultaneously. Second, both
$\|\widehat f\|_1$ and $s(f)$ are non-increasing under affine restriction
(Lemma~\ref{lem:restriction}); we may therefore use the \emph{original} value of
$\|\widehat f\|_1$ at every node, and the logarithms never compound. This is the reason the
final bound is linear, rather than polynomial, in $\log s$.

The entire difficulty is thus concentrated in a single step: bounding the codimension
needed to kill one degree. We denote by
\[
  T \;=\; \Delta^{d} p, \qquad \Delta_h p(x)=p(x+h)+p(x),
\]
the $d$-th finite-difference form of a degree-$d$ polynomial; it is a symmetric
$d$-linear form which vanishes whenever two of its arguments coincide
(Lemma~\ref{lem:Tj}). Our bound on the required codimension is assembled from four
lemmas.

\paragraph{Step 1: binary contraction ranks (Subsection~\ref{sec:step1}).}
Let $M(T)$ be the largest rank over $\F_2$ of $T(h_1,\dots,h_{d-2},-,-)$ at binary
anchors. Taking $d-2$ differences reduces $p$ to degree at most two; Dickson's theorem
and submultiplicativity of the spectral norm give
\[
  M(T)\le2^{d-1}\log_2\|\widehat f\|_1.
\]

\paragraph{Step 2: generic versus binary rank (Subsection~\ref{sec:step2}).}
For $d\ge3$, let $K=\F_2(\mathbf t_1,\dots,\mathbf t_{d-2})$, with independent formal
anchor blocks of length $n$, and let $N$ be the matrix of
$T(\mathbf t_1,\dots,\mathbf t_{d-2},-,-)$. Since its entries are linear and homogeneous
in each block, the rank comparison of Moshkovitz and Zhu~\cite[Theorem~1.1]{MoshkovitzZhu}
gives
\[
  R:=\rank_KN\le2^{d-2}M(T).
\]

\paragraph{Step 3: a strong vanishing identity (Subsection~\ref{sec:step3}).}
A perturbation and Schur-complement argument shows that the generic contraction
$A=T(\mathbf t_1,\dots,\mathbf t_{d-2},-,-)$ has a radical $U=\ker A\le K^n$ satisfying
\[
  \dim_K U=n-R,\qquad T(K^n,U,\dots,U)=0.
\]

\paragraph{Step 4: descent to $\F_2$ (Subsection~\ref{sec:step4}).}
Successive normalised specialisations produce $U'\le\F_2^n$ with
\[
  \dim_{\F_2}U'=\dim_KU=n-R,\qquad T(\F_2^n,U',\dots,U')=0.
\]
Thus $T(U',\dots,U')=0$, giving a degree-reducing subspace of codimension $R$.

\paragraph{Assembling.}
At a node whose current phase $p$ has degree $j\ge3$, the above gives a subspace (in $\F_2^n$) of codimension at most $2^{j-2}M(\Delta^j p)\le2^{2j-3}\log_2\|\widehat f\|_1$ on which $\Delta^j p$ vanishes. 
At $j=2$, Dickson's theorem provides a totally isotropic subspace of codimension $r\le\log_2\|\widehat f\|_1$, after which one further query suffices. Summing these up, we have
\[
  \Dxor(f)\;\le\;1+\Bigl(1+\sum_{j=3}^{d}2^{2j-3}\Bigr)\log_2\|\widehat f\|_1
  \;=\;1+\frac{2^{2d-1}-5}{3}\log_2\|\widehat f\|_1 .
\]

\subsection{Organisation}

Section~\ref{sec:prelim} fixes notation and collects the facts we use, with proofs
of the ones that are short. Section~\ref{sec:main-proof} contains the proof of the main
theorem: Subsection~\ref{sec:overview} states the degree-reduction scheme,
Subsections~\ref{sec:step1}--\ref{sec:step4} prove the four steps, and
Subsection~\ref{sec:proof} assembles them into Theorems~\ref{thm:main}
and~\ref{thm:main-weight} and delimits the range in which the bound is informative.

\section{Preliminaries}\label{sec:prelim}

All logarithms are base two unless indicated otherwise. We write $[n]=\{1,\dots,n\}$ and
$e_1,\dots,e_n$ for the standard basis of $\F_2^n$. Unless another base field is specified,
$V$ denotes an $\F_2$-vector space of the ambient dimension under discussion.

\subsection{Algebraic notation and basic facts}

We review the algebraic conventions needed in Subsections~\ref{sec:step2}--\ref{sec:step4}.
Throughout this review, $\F$ is a field and $x=(x_1,\dots,x_m)$ is a tuple of independent
formal variables. In characteristic two, $1+1=0$; a field of characteristic two need not
be $\F_2$ itself.

\paragraph{Polynomial rings, function fields, and evaluation.}
\key{The \emph{polynomial ring} $\F[x_1,\dots,x_m]$, abbreviated $\F[x]$, consists of finite sums of monomials with coefficients in $\F$. Equality means equality of all coefficients.}
\key{The \emph{total degree} of a monomial $x_1^{d_1}\cdots x_m^{d_m}$ is
$d_1+\cdots+d_m$. The total degree $\deg P$ of a nonzero polynomial $P$ is the
maximum total degree of its monomials with nonzero coefficients.}
It has no zero divisors, but its only invertible elements are the nonzero constants.
\key{Its \emph{rational function field} is the field of fractions
\[
  \F(x_1,\dots,x_m)=\left\{\frac{p(x)}{q(x)}:
  p,q\in\F[x_1,\dots,x_m],\ q\neq0\right\}.
\]
Two fractions $p/q$ and $p'/q'$ are equal when $pq'=p'q$.} Note that square brackets denote the polynomial ring; parentheses denote the field obtained by allowing division by every
nonzero polynomial. 
Throughout this paper, we work with \key{rational function fields generated by formal variables that are \emph{algebraically independent} over $\F$. These fields are \emph{purely
transcendental extensions} of $\F$.}

For $a\in\F^m$, substitution $x=a$ defines a ring homomorphism
$\F[x]\to\F$, so it preserves sums, products, and polynomial identities. A rational
function can be evaluated at $a$ only if it admits a representation $p/q$ with $q(a)\neq0$;
it is then called \emph{regular at $a$}. For example, $x/x=1$ is regular at zero after
cancellation, whereas $1/x$ is not. Formal nonvanishing is different from nonvanishing
on the binary cube: $x^2+x$ is nonzero in $\F_2[x]$ but vanishes at both $0$ and $1$.
In particular, $x_i^2=x_i$ is not an identity in $\F_2[x]$; it holds only after binary
evaluation. The multilinear representatives of Boolean functions are discussed below.

\paragraph{Vector spaces and extension of scalars.}
For a field extension $E/\F$, we regard $\F^m$ as a subset of $E^m$ and allow vector coordinates and linear-combination coefficients to lie in $E$. 
\key{Dimension is always relative to the scalar field: $\dim_E W$ counts basis vectors when linear combinations use coefficients in $E$.} 
We write $U\le E^m$ to mean that $U$ is an $E$-linear
subspace of $E^m$; unless otherwise stated, its dimension is taken over $E$.
The standard basis of $E^m$ is $e_1,\dots,e_m$, so
\[
  \dim_E E^m=m,\qquad \codim_{E^m}U=m-\dim_E U.
\]
The number of formal variables generating $E$ does not affect this dimension. 
\explain{For example, $\F_2(t)^m$ has dimension $m$ over $\F_2(t)$, but is infinite-dimensional over $\F_2$ when $m\ge1$: the vectors $e_1,te_1,t^2e_1,\dots$ are linearly independent over $\F_2$.}

Extending scalars is different from viewing the same space over a smaller field. If
$W\le\F^m$ and $W_E=\operatorname{span}_E W$, then $\dim_E W_E=\dim_{\F}W$.
A multilinear form extends by using its original coefficients and allowing its arguments
to lie in $E^m$; polynomial identities remain valid under this extension. Similarly,
``without losing dimension'' in Subsection~\ref{sec:step4} means
$\dim_K U=\dim_{\F_2}U'=k$: the two dimensions are measured over their respective
scalar fields, not both over $\F_2$.

\paragraph{Matrix rank, minors, and specialisation.}
{ }
\key{For a matrix $N$ over a field $E$, $\rank_E N$ is the dimension over $E$ of its column space, equivalently its row space or the largest order of a nonzero minor.} 
A square matrix is invertible exactly when its determinant is nonzero. The rank of a matrix over
$\F$ does not change on extending scalars to $E$, since its minors do not change.
\key{For a polynomial matrix $N(x)$, its \emph{generic rank} is instead computed over $K=\F(x)$, where polynomial entries may be divided by nonzero polynomials. It is the largest size of a minor whose determinant is not the zero polynomial. Specialisation
can lower this rank but cannot increase it:}
\[
  \key{\rank_{\F}N(a)\le\rank_{\F(x)}N(x).}
\]
\explain{Indeed, all minors larger than the generic rank vanish identically, hence vanish at every $a$. For example, $\operatorname{diag}(x,x+1)$ has rank two over $\F_2(x)$, but rank one over $\F_2$ when $x=0$ or $x=1$.} Generic and binary evaluated ranks are defined
for variable blocks in Subsection~\ref{sec:step2}.

\paragraph{Bilinear forms, radicals, and non-degeneracy.}
For a \key{symmetric bilinear form} $A:E^m\times E^m\to E$, its \key{matrix} in the standard basis
is defined by $N_{rs}=A(e_r,e_s)$, so $A(u,v)=u^{\mathsf T}Nv$; we write $\rank_E A=\rank_E N$.
Its \key{\emph{radical}, also denoted $\ker A$}, is
\[
  \operatorname{rad}A=\{u\in E^m:A(u,v)=0\text{ for every }v\in E^m\}
  =\ker N.
\]
Rank--nullity gives $\codim_{E^m}\operatorname{rad}A=\rank_E A$. A form is
\emph{non-degenerate} when its radical is zero. \key{A subspace $W$ is \emph{totally isotropic}
when $A(W,W)=0$}; this does not require $W$ to lie in the radical, whose vectors pair to
zero with the \emph{whole} space. If $U=\operatorname{rad}A$ and $E^m=S\oplus U$, the
restriction to $S$ is non-degenerate and the matrix in an adapted basis has the form
$\left(\begin{smallmatrix}P_0&0\\0&0\end{smallmatrix}\right)$ with $P_0$ invertible.

\paragraph{Schur complements and formal power series.}
Over any field $E$, a block matrix with invertible square block $P$ satisfies
\[
  \key{\rank_E\begin{pmatrix}P&B\\C&H\end{pmatrix}
  =\rank_E P+\rank_E(H-CP^{-1}B)}.
\]
The second term is the rank of the \key{\emph{Schur complement} $H-CP^{-1}B$}; the identity follows by
invertible block row and column operations. In particular, if the full rank is at most
$\rank_E P$, that complement is zero.

\key{Double brackets denote the \emph{formal power series ring}}
\[
  \key{E[[t_1,\dots,t_\ell]]
  =\left\{\sum_{\mathbf d\in\Z_{\ge0}^{\ell}}c_{\mathbf d}t^{\mathbf d}:
  c_{\mathbf d}\in E\right\},\qquad t^{\mathbf d}=\prod_i t_i^{d_i}.}
\]
Unlike the polynomial ring $E[t_1,\dots,t_\ell]$, this ring allows infinitely many nonzero
coefficients.
All exponents are nonnegative.
Two series are equal exactly when all corresponding coefficients agree. No analytic
convergence or numerical assignment to the $t_i$ is required. Multiplication is
well-defined because the coefficient of $t^{\mathbf d}$ in a product is
\[
  \sum_{\mathbf e+\mathbf f=\mathbf d}c_{\mathbf e}b_{\mathbf f},
\]
a finite sum for each fixed $\mathbf d$. For example,
\[
  \key{(1-t)^{-1}=1+t+t^2+\cdots\qquad\text{in }E[[t]]},
\]
\key{since multiplying the right-hand side by $1-t$ gives $1$ coefficient by coefficient};
no condition such as $|t|<1$ is involved, even in characteristic two.

\key{Let $\mathfrak m=(t_1,\dots,t_\ell)$ denote the ideal generated by $t_1,\dots,t_\ell$ in $E[[t_1,\dots,t_\ell]]$. It consists precisely of series with zero constant term.
Its $r$-th power $\mathfrak m^r$ consists of series with no terms of total degree below $r$.}
A series is invertible exactly when its constant term is nonzero, and a matrix of series is invertible exactly when its constant matrix is invertible. For a matrix $J(t)$ with
zero constant matrix,
\[
  (I-J(t))^{-1}=\sum_{r\ge0}J(t)^r.
\]
Every term in $J(t)^r$ has total degree at least $r$, so a coefficient of total degree
$D$ receives contributions only from $r\le D$. This makes the infinite sum formally
well-defined. Rational functions regular at the origin have unique such expansions;
$1/t_1$, by contrast, has a pole and does not belong to this series ring.


\paragraph{Valuations and residue at zero.}
\key{In one variable, let $\operatorname{ord}_0p$ be the smallest exponent with nonzero coefficient in a nonzero polynomial $p(t)$. Equivalently, $p(t)=t^r u(t)$ with $u(0)\neq0$, and $\operatorname{ord}_0p=r$.} 
\explain{For example, $\operatorname{ord}_0(1+t^2)=0$ and $\operatorname{ord}_0(t^2+t^3)=2$. Note that this is the order of vanishing at zero, not the polynomial's degree; it is defined algebraically, without an analytic notion of a neighbourhood.}
\key{The \emph{$t$-adic valuation} on $E(t)$ is}
\[
  \key{\nu_t(p/q)=\operatorname{ord}_0p-\operatorname{ord}_0q,
  \qquad \nu_t(0)=\infty.}
\]
\explain{This extends the polynomial order of vanishing to rational functions by subtracting
the denominator's order from the numerator's. A positive valuation records the order
of a zero, and a negative valuation records the negative of the order of a pole. For example,
$\nu_t(t^2/(1+t)) = 2-0 = 2$ and $\nu_t(1/t)=0-1 = -1$.} The valuation satisfies $\nu_t(fg)=\nu_t(f)+\nu_t(g)$ and $\nu_t(f+g)\ge\min\{\nu_t(f),\nu_t(g)\}$.
The \emph{valuation ring} $\mathcal O=\{f\in E(t):\nu_t(f)\ge0\}$ is precisely the ring of rational functions regular at zero.
\explain{For nonzero $f=p/q$, write $p=t^a p_0$ and $q=t^b q_0$ with
$p_0(0),q_0(0)\neq0$. Then $f=t^{a-b}p_0/q_0$, so evaluation at zero is possible exactly when $\nu_t(f)=a-b\ge0$: after cancellation, the denominator is nonzero at zero.}
The original denominator may vanish there, as in $t/t=1$, without preventing evaluation of the rational function. Negative valuation leaves a power of $t$ in the denominator and hence a pole. The zero function also evaluates to zero.
\key{The resulting \emph{residue map} $\mathcal O\to E$, $f\mapsto f|_{t=0}$, is a ring homomorphism.
The residue map preserves identities, but not necessarily matrix rank}:
$\operatorname{diag}(1,t)$ has rank 2 over $E(t)$, but its value
$\operatorname{diag}(1,0)$ has rank 1 over $E$.
However, \key{if a matrix over $\mathcal O$ has $k$ columns and a $k\times k$ identity submatrix, that submatrix survives evaluation and guarantees rank $k$ afterwards.} This is the
dimension-preserving normalisation used in Subsection~\ref{sec:step4}.

\subsection{Fourier analysis on the cube}

We use $f$ for a real-valued function on $\F_2^n$, and $f,g$ when two functions are
needed. When a function is sign-valued, we state this explicitly; for a polynomial
$p\colon\F_2^n\to\F_2$, its associated sign function is $f=(-1)^p$.
For $\xi\in\F_2^n$ let $\chi_\xi(x)=(-1)^{\xi\cdot x}$, where
$\xi\cdot x=\sum_i\xi_ix_i\bmod2$. The characters $\{\chi_\xi\}$ form an orthonormal
basis of the space of real functions on $\F_2^n$ under the inner product
$\langle f,g\rangle=\Ex_x[f(x)g(x)]$. Every $f\colon\F_2^n\to\R$ has the unique expansion
$f=\sum_\xi\widehat f(\xi)\chi_\xi$ with
$\widehat f(\xi)=\Ex_x[f(x)\chi_\xi(x)]$. We set
\[
  \supp\widehat f=\{\xi:\widehat f(\xi)\neq0\},\qquad
  s(f)=|\supp\widehat f|,\qquad
  \|\widehat f\|_1=\sum_\xi|\widehat f(\xi)| .
\]
For $f$ taking values in $\{-1,1\}$, Parseval's identity gives
$\sum_\xi\widehat f(\xi)^{2}=1$, whence by Cauchy--Schwarz
\begin{equation}\label{eq:parseval}
  1\;\le\;\|\widehat f\|_1\;\le\;\sqrt{s(f)} .
\end{equation}
Equality on the left holds exactly when $f=\pm\chi_\xi$ for some $\xi$, i.e.\ when $s=1$.

The following two facts are standard, and can be found in, for example, \cite{TWXZ13}.
\begin{fact}[Convolution and products~\cite{TWXZ13}]\label{fact:product}
For $f,g\colon\F_2^n\to\R$, $\widehat{fg}=\widehat f*\widehat g$, hence
$\|\widehat{fg}\|_1\le\|\widehat f\|_1\|\widehat g\|_1$. Moreover
$\|\widehat{f_a}\|_1=\|\widehat f\|_1$ for the translate $f_a(x)=f(x+a)$, and
$\|\widehat{\chi_\eta f}\|_1=\|\widehat f\|_1$.
\end{fact}


\begin{lemma}[Affine restriction~\cite{TWXZ13}]\label{lem:restriction}
Let $f\colon\F_2^n\to\R$, let $H=x_0+V'$ be an affine subspace of $\F_2^n$, and let
$f|_H$ denote the restriction of $f$ to $H$, regarded as a function on
$\F_2^{\dim V'}$ via a choice of basis of $V'$. Then
\[
  s(f|_H)\le s(f),\qquad \|\widehat{f|_H}\|_1\le\|\widehat f\|_1 .
\]
If the affine map is invertible (i.e.\ $H=\F_2^n$), both quantities are preserved;
when $f$ is sign-valued, $\Dxor(f)$ is preserved as well.
\end{lemma}


\begin{definition}[Granularity]
For a sign-valued function $f\colon\F_2^n\to\{-1,1\}$, its \emph{granularity} $G(f)$
is the least integer $G\ge0$ with $\widehat f(\xi)\in2^{-G}\Z$ for all $\xi$.
\end{definition}

\begin{lemma}[Granularity lower bound~\cite{ChistopolskayaPodolskii}]\label{lem:granularity}
For nonconstant $f\colon\F_2^n\to\{-1,1\}$ one has $\Dxor(f)\ge G(f)+1$.
\end{lemma}



\subsection{Parity decision trees}

\begin{definition}
A \emph{parity decision tree} over $\F_2^n$ is a rooted binary tree in which every
internal node is labelled by a vector $a\in\F_2^n\setminus\{0\}$ and every leaf by a
value in $\{-1,1\}$. On input $x$ one starts at the root and, at a node labelled $a$,
descends to the child indexed by $a\cdot x$. The tree \emph{computes} $f$ if the leaf
reached always carries the value $f(x)$. $\Dxor(f)$ is the minimum depth of such a tree.
\end{definition}

The set of inputs reaching a node at depth $c$ is an affine subspace of codimension at
most $c$, and the two children partition it into two parallel cosets. Consequently a PDT
of depth $D$ induces a partition of $\F_2^n$ into at most $2^{D}$ affine subspaces on each
of which $f$ is constant.

\begin{remark}[Affine queries are free]\label{rem:affine-queries}
Our construction works on an affine subspace $H=x_0+V'$ and produces linear forms in the
coordinates of $V'$. Pulled back to $\F_2^n$ such a form becomes an affine function
$\xi\cdot x+c$ with $c\in\F_2$ a constant known to the algorithm. Querying $\xi\cdot x$
and flipping the branch label implements it at the same cost, so we may freely speak of
querying affine forms.
\end{remark}

We also use the following repeatedly. If we query a basis of
$U^{\perp}=\{a: a\cdot u=0\ \forall u\in U\}$ for a subspace $U\le V'$, then we spend
exactly $\codim U$ queries and the residual input set at each of the $2^{\codim U}$ leaves
is a coset of $U$.

\subsection{\texorpdfstring{$\F_2$}{F2}-polynomials, finite differences, and alternating forms}

Every $p\colon\F_2^n\to\F_2$ has a unique \emph{algebraic normal form}
$p(x)=\sum_{S\subseteq[n]}c_S\prod_{i\in S}x_i$ with $c_S\in\F_2$; we write
$\deg p=\max\{|S|:c_S\neq0\}$. Möbius inversion gives
\begin{equation}\label{eq:mobius}
  c_S\;=\;\sum_{x\,:\,\supp(x)\subseteq S} p(x) \pmod 2 .
\end{equation}

For $h\in\F_2^n$ let $\Delta_h p(x)=p(x+h)+p(x)$. \key{The operators $\Delta_h$ commute} and
satisfy $\deg\Delta_hp\le\deg p-1$ and
\begin{equation}\label{eq:delta-additive}
  \key{\Delta_{h+h'}=\Delta_h+\Delta_{h'}+\Delta_h\Delta_{h'}}.
\end{equation}
Indeed, both sides applied to $p$ at $x$ equal $p(x+h+h')+p(x)$.

The next lemma describes the iterated finite differences obtained by differentiating
a polynomial $p$ exactly $d=\deg p$ times.

\begin{lemma}\label{lem:Tj}
Let $\deg p=d$. Then
\[
  T(h_1,\dots,h_d)\;:=\;\Delta_{h_1}\cdots\Delta_{h_d}p(x)
\]
is independent of $x$, and as a function of $(h_1,\dots,h_d)$ it is
\begin{enumerate}[label=(\roman*),itemsep=1pt]
  \item \emph{symmetric}, and
  \item \emph{multilinear}, and
  \item \emph{alternating} in the strong sense that $T=0$ whenever two of its
        arguments are equal.
\end{enumerate}
Writing $p_d=\sum_{|S|=d}c_S\prod_{i\in S}x_i$ for the degree-$d$ homogeneous part
of $p$, we have
\begin{equation}\label{eq:T-on-basis}
  T(e_{i_1},\dots,e_{i_d})\;=\;c_{\{i_1,\dots,i_d\}}
  \quad\text{for distinct }i_1,\dots,i_d .
\end{equation}
Finally, $T\neq0$.
\end{lemma}

\begin{proof}
Applying $d+1$ difference operators to a polynomial of degree $d$ gives $0$, so
$\Delta_h(T)=0$ for every $h$, i.e.\ $T$ is constant in $x$. Symmetry is commutativity
of the $\Delta$'s.

For (iii) it suffices, by symmetry, to treat two adjacent equal arguments, and then
$\Delta_h\Delta_h=0$ as an operator: for any $p\colon\F_2^n\to\F_2$,
\[
  \Delta_h\Delta_hp(x)=\bigl(p(x+h+h)+p(x+h)\bigr)+\bigl(p(x+h)+p(x)\bigr)
  =p(x)+p(x)=0,
\]
using $h+h=0$ in $\F_2^n$ and cancellation in $\F_2$.

For (ii), by \eqref{eq:delta-additive},
\[
  T(h_1+h_1',h_2,\dots)=T(h_1,h_2,\dots)+T(h_1',h_2,\dots)
   +\Delta_{h_1}\Delta_{h_1'}\Delta_{h_2}\cdots\Delta_{h_d}p ,
\]
and the last term applies $2+(d-1)=d+1$ operators to $p$, hence vanishes.

Identity \eqref{eq:T-on-basis} follows from $\Delta_{e_i}\bigl(\prod_{\ell\in S}x_\ell\bigr)
=\prod_{\ell\in S\setminus\{i\}}x_\ell$ if $i\in S$ and $0$ otherwise: applying
$\Delta_{e_{i_1}}\cdots\Delta_{e_{i_d}}$ to a monomial $x_S$ with $|S|\le d$ gives a
nonzero (indeed constant $1$) result iff $S=\{i_1,\dots,i_d\}$. Since $\deg p=d$, some
$c_S\neq0$ with $|S|=d$, so $T\neq0$.
\end{proof}

By \eqref{eq:T-on-basis}, $T$ records exactly the degree-$d$ part of $p$, and the lower-order terms of $p$ are invisible to it. \key{We refer to $T$ as the \emph{top form} of $p$}.
We write \key{$\Delta^d p:=T$} for this entire $d$-linear form, with the anchors as its arguments. 
Thus $(\Delta^d p)(h_1,\ldots,h_d)=\Delta_{h_1}\cdots\Delta_{h_d}p(x)$, independently of $x$. 

The following lemma says that if a polynomial $p$ over $\F_2$ has a strictly lower degree when restricted on a subspace $U$, then so does it when restricted on every coset of $U$. This fact was established in \cite{TWXZ13}, and we rewrite this in the language of top forms and include a short proof for completeness.

\begin{lemma}[Degree reduction criterion]\label{lem:degree-drop}
Let $\deg p=d$ with top form $T=\Delta^d p$, and let $U\le\F_2^n$ be a subspace.
The following are equivalent:
\begin{enumerate}[label=(\alph*),itemsep=1pt]
  \item $T(u_1,\dots,u_d)=0$ for all $u_1,\dots,u_d\in U$;
  \item $\deg\bigl(p|_{a+U}\bigr)\le d-1$ for every $a\in\F_2^n$.
\end{enumerate}
\end{lemma}

\begin{proof}
If $\dim U<d$ then (a) holds automatically (any $d$ vectors of $U$ are linearly dependent,
and a multilinear alternating form vanishes on dependent tuples), and so does (b), since a
polynomial on fewer than $d$ variables has degree $<d$. So assume $\dim U=\mu\ge d$.

Fix a basis $u_1,\dots,u_\mu$ of $U$ and set
$q(y)=p\bigl(a+\sum_{i=1}^{\mu}y_iu_i\bigr)$, a polynomial on $\F_2^{\mu}$ of
degree at most $d$. The key computation is
\begin{equation}\label{eq:pullback-delta}
  \Delta^{(y)}_{e_i}q\,(y)\;=\;q(y+e_i)+q(y)
  \;=\;p\Bigl(a+\textstyle\sum_\ell y_\ell u_\ell+u_i\Bigr)+p\Bigl(a+\textstyle\sum_\ell y_\ell u_\ell\Bigr)
  \;=\;\bigl(\Delta_{u_i}p\bigr)\Bigl(a+\textstyle\sum_\ell y_\ell u_\ell\Bigr),
\end{equation}
i.e.\ differencing $q$ along $e_i$ corresponds to differencing $p$ along $u_i$.
Iterating \eqref{eq:pullback-delta} for distinct indices $i_1,\dots,i_d$, the
coefficient of $\prod_{\ell=1}^{d}y_{i_\ell}$ in $q$ equals
\[
  \Delta^{(y)}_{e_{i_1}}\cdots\Delta^{(y)}_{e_{i_d}}q
  \;=\;\Delta_{u_{i_1}}\cdots\Delta_{u_{i_d}}p
  \;=\;T(u_{i_1},\dots,u_{i_d}).
\]
For $\deg q=d$, the coefficient identity follows from Lemma~\ref{lem:Tj} and
\eqref{eq:T-on-basis}; for $\deg q<d$, both the degree-$d$ coefficient and the
$d$-fold difference of $q$ are zero. In either case, the coefficient does not
depend on the offset $a$.

Hence all degree-$d$ coefficients of $q$ vanish, for every $a$, if and only if
$T$ vanishes on all $d$-tuples of \emph{distinct} basis vectors of $U$. By multilinearity
this is equivalent to (a): expanding arbitrary $u_1,\dots,u_d\in U$ in the basis, every
term involving a repeated basis vector vanishes by the alternating property
(Lemma~\ref{lem:Tj}(iii)), and the remaining terms are exactly the values on distinct
basis vectors.
\end{proof}


The following fact relates the degree of a Boolean function to its sparsity. (The $\|\widehat f\|_1>1$ part is due to the equality
case of \eqref{eq:parseval}.)
\begin{lemma}[Bernasconi--Codenotti~\cite{BC99}]\label{lem:d-le-logs}
If $f=(-1)^{p}$ with $\deg p=d\ge2$, then $s(f)\ge2^{d}$; in particular
$d\le\log_2 s(f)$ and $\|\widehat f\|_1>1$.
\end{lemma}


\paragraph{The sign-valued formulation.}
For sign-valued functions, the same finite-difference operation takes a multiplicative
form.

\begin{definition}
For $f\colon\F_2^n\to\{-1,1\}$ and $h\in\F_2^n$ set $D_hf(x)=f(x)f(x+h)$.
\end{definition}

If $f=(-1)^{p}$ then $D_hf=(-1)^{\Delta_hp}$, so the multiplicative derivative of the
sign corresponds to the additive difference of the phase. Iterating,
\begin{equation}\label{eq:iterated-deriv}
  D_{h_1}\cdots D_{h_k}f(x)\;=\;\prod_{\eps\in\{0,1\}^{k}}f\Bigl(x+\sum_{i=1}^{k}\eps_ih_i\Bigr)
  \;=\;(-1)^{\Delta_{h_1}\cdots\Delta_{h_k}p(x)} ,
\end{equation}
a product of $2^{k}$ translates of $f$. Combining with Fact~\ref{fact:product}:

\begin{lemma}\label{lem:deriv-norm}
$\bigl\|\widehat{D_{h_1}\cdots D_{h_k}f}\bigr\|_1\le\|\widehat f\|_1^{\,2^{k}}$.
\end{lemma}

\subsection{Quadratic functions and Dickson's theorem}

Let $p\colon\F_2^n\to\F_2$ have degree at most $2$. Its \emph{polar form} is
\[
  \beta_p(u,v) \;:= \Delta_u\Delta_vp \;=\;p(u+v)+p(u)+p(v)+p(0)\;,
\]
For $\deg p=2$, the second equality follows from Lemma~\ref{lem:Tj} by evaluating
at $0$; for $\deg p<2$, the second difference is identically zero, so both sides
are zero. In either case, the polar form is an alternating bilinear form.
Over $\F_2$ such a form has even rank, say $2r$, and
Dickson's theorem (see e.g.~\cite[Ch.~15]{MacWilliamsSloane}) provides an invertible
linear change of coordinates after which
\begin{equation}\label{eq:dickson}
  p(y)\;=\;\sum_{i=1}^{r}y_{2i-1}y_{2i}\;+\;\ell(y)\;+\;c
\end{equation}
for a linear form $\ell$ and $c\in\F_2$. The following lemma collects basic properties of quadratic polynomials.

\begin{lemma}\label{lem:quadratic}
Let $p$ have polar rank $2r$ and let $f=(-1)^{p}$. Then
\[
  s(f)=2^{2r},\qquad
  |\widehat f(\xi)|\in\{0,2^{-r}\},\qquad
  \|\widehat f\|_1=2^{r},\qquad G(f)=r .
\]
Moreover $\supp\widehat f$ is a coset of a $2r$-dimensional subspace.
\end{lemma}


\begin{proposition}[The quadratic case is exact]\label{prop:quadratic-exact}
If $\deg p=2$ with polar rank $2r\ge2$ and $f=(-1)^{p}$, then
\[
  \Dxor(f)\;=\;r+1\;=\;1+\half\log_2 s(f).
\]
\end{proposition}

\begin{proof}
Upper bound: after the change of coordinates \eqref{eq:dickson} query
$y_1,y_3,\dots,y_{2r-1}$ ($r$ queries); on each branch the residual phase is affine, so
one further query determines $f$ (Remark~\ref{rem:affine-queries}). Lower bound:
$G(f)=r$ by Lemma~\ref{lem:quadratic} and $\Dxor(f)\ge G(f)+1$ by
Lemma~\ref{lem:granularity}. Finally $s=2^{2r}$.
\end{proof}

\subsection{Threshold weight}

\begin{definition}
A \emph{margin-one sign witness} for $f\colon\F_2^n\to\{-1,1\}$ is a real function $g$ with
$f(x)g(x)\ge1$ for all $x$. The \emph{threshold weight} is
$W(f)=\min\{\|\widehat g\|_1: g\text{ a margin-one sign witness}\}$.
\end{definition}

Taking $g=f$ gives $W(f)\le\|\widehat f\|_1\le\sqrt{s(f)}$. We record the properties
we need.

\begin{lemma}\label{lem:weight}
\begin{enumerate}[label=(\alph*),itemsep=1pt]
  \item If $g$ witnesses $f$ then $g|_H$ witnesses $f|_H$ for every affine $H$, and
        $\|\widehat{g|_H}\|_1\le\|\widehat g\|_1$. Hence $W(f|_H)\le W(f)$.
  \item If $g_i$ witnesses $f_i$ for $i=1,2$ then $g_1g_2$ witnesses $f_1f_2$, and
        $\|\widehat{g_1g_2}\|_1\le\|\widehat{g_1}\|_1\|\widehat{g_2}\|_1$. In particular,
        by \eqref{eq:iterated-deriv},
        $W(D_{h_1}\cdots D_{h_k}f)\le W(f)^{2^{k}}$.
  \item If $p$ has degree at most $2$ and polar rank $2r$, let $f=(-1)^p$.
        Every margin-one sign witness $g$ for $f$ satisfies $\|\widehat g\|_1\ge2^{r}$;
        hence $W(f)=2^{r}$.
  \item If $\deg p=d\ge2$ and $f=(-1)^{p}$ then $W(f)>1$.
\end{enumerate}
\end{lemma}

\begin{proof}
(a) and (b) are immediate from the definitions and Fact~\ref{fact:product} (for (a), use
Lemma~\ref{lem:restriction} applied to $g$). For (c), let $f=(-1)^{p}$ and let $g$
be a margin-one sign witness for $f$. Then
\[
  1\;\le\;\Ex_x[f(x)g(x)]\;=\;\sum_\xi\widehat f(\xi)\widehat g(\xi)
  \;\le\;\max_\xi|\widehat f(\xi)|\cdot\|\widehat g\|_1\;=\;2^{-r}\|\widehat g\|_1 ,
\]
using $fg\ge1$ pointwise, Plancherel, and Lemma~\ref{lem:quadratic}. The matching upper
bound is $W(f)\le\|\widehat f\|_1=2^{r}$. For (d), suppose $W(f)=1$ with witness $g$. Then
$\|g\|_\infty\le\|\widehat g\|_1=1$, while $fg\ge1$ forces $|g|\ge1$; hence $|g|\equiv1$
and $g=f$. So $\|\widehat f\|_1=1$, i.e.\ $s(f)=1$, contradicting
Lemma~\ref{lem:d-le-logs}.
\end{proof}

Because of Lemma~\ref{lem:weight} every occurrence of $\|\widehat f\|_1$ in the proof below
may be replaced by $W(f)$; we will carry a single symbol
\[
  B\;=\;B(f)\;\in\;\bigl\{\|\widehat f\|_1,\;W(f)\bigr\}
\]
and only use that (i) $B$ does not increase under affine restriction, (ii) $B$ is
submultiplicative under the products in \eqref{eq:iterated-deriv}, (iii) a quadratic
phase of polar rank $2r$ has $B=2^{r}$, and (iv) $\log_2B(f)>0$ when $f=(-1)^p$ with $\deg p\ge2$
(Lemmas~\ref{lem:d-le-logs} and~\ref{lem:weight}(d)).

\section{Degree reduction and proof of the main theorem}\label{sec:main-proof}

This section proves Theorems~\ref{thm:main} and~\ref{thm:main-weight}.
Subsection~\ref{sec:overview} sets up the degree-reduction framework, and
Subsections~\ref{sec:step1}--\ref{sec:step4} establish its four steps.
Finally, Subsection~\ref{sec:proof} combines these ingredients to complete the proofs.

\subsection{The degree-reduction scheme}\label{sec:overview}

\begin{definition}[Degree-reduction cost]
For a symmetric $d$-linear form $T$ on $\F_2^{m}$ which vanishes on tuples with a repeated
argument, let
\[
  c(T)\;=\;\min\bigl\{\codim U : U\le\F_2^{m},\ T(U,\dots,U)=0\bigr\} .
\]
\end{definition}

For a polynomial $p\colon\F_2^m\to\F_2$ of degree exactly $d$, let
$T=\Delta^d p$ be its top form as in Lemma~\ref{lem:Tj}.
By Lemma~\ref{lem:degree-drop}, $c(\Delta^d p)=c(T)$ is the minimum size of a fixed
batch of parity queries that lowers the degree to at most $d-1$ on every branch.
A minimising subspace $U$ realises this cost by querying a basis of $U^\perp$.
The condition $T(U^d)=0$ quantifies over all anchor tuples in $U^d$.
The following lemma formalises the recursion, with $j$ indexing the degree levels
up to $d$.

\begin{lemma}[Recursion]\label{lem:recursion}
Let $p\colon\F_2^n\to\F_2$ have $\deg p = d$, and $f=(-1)^p$. Suppose that for some function
$\Gamma\colon\{2,3,\dots\}\to\R_{\ge0}$ the following holds:
for every $3\le j\le d$ and every polynomial $q$ of degree exactly $j$ arising as an affine
restriction of $p$, one has $c(\Delta^{j}q)\le\Gamma(j)$; and every
degree-$2$ affine restriction $q$ of $p$ satisfies $\Dxor((-1)^{q})\le\Gamma(2)+1$. Then
\[
  \Dxor(f)\;\le\;1+\sum_{j=2}^{d}\Gamma(j).
\]
\end{lemma}

\begin{proof}
Induct on $d$. If $d\le1$ then $f=\pm\chi_\xi$ and one query suffices (zero if $f$
is constant), which is at most $1+\sum_{j=2}^{d}\Gamma(j)$ as the sum is empty. If $d=2$
the hypothesis gives $\Dxor\le\Gamma(2)+1$. If $d\ge3$, pick $U$ with
$\codim U\le\Gamma(d)$ and $\Delta^{d}p(U^{d})=0$, and query a basis of $U^{\perp}$. By
Lemma~\ref{lem:degree-drop} the restriction of $p$ to each of the resulting cosets has
degree at most $d-1$; each such restriction is an affine restriction of $p$, and an affine
restriction of an affine restriction is again one, so the inductive hypothesis applies with
the \emph{same} $\Gamma$ and gives depth at most $1+\sum_{j=2}^{d-1}\Gamma(j)$ below each
leaf. The total depth is at most $\Gamma(d)+1+\sum_{j=2}^{d-1}\Gamma(j)$.
\end{proof}


Our task is therefore reduced to a single inequality for $\deg p=d\ge3$ and $f=(-1)^p$:
\begin{equation}\label{eq:task}
  c(\Delta^{d}p)\;\le\;2^{2d-3}\log_2 B(f).
\end{equation}
This occupies Subsections~\ref{sec:step1}--\ref{sec:step4}.

\subsection{Step 1: contraction ranks and the spectral norm}\label{sec:step1}

\begin{definition}
Let $T$ be a symmetric $d$-linear form on $\F_2^{n}$, $d\ge2$. For
$h_1,\dots,h_{d-2}\in\F_2^{n}$ the \emph{two-slot contraction}
$T(h_1,\dots,h_{d-2},-,-)$ is a bilinear form; let
\[
  M(T)\;=\;\max_{h_1,\dots,h_{d-2}\in\F_2^{n}}\rank_{\F_2}T(h_1,\dots,h_{d-2},-,-).
\]
For $d=2$ there are no anchors to fix, and $M(T)=\rank_{\F_2}T$.
\end{definition}

\begin{lemma}[Step 1]\label{lem:step1}
Let $p$ on $\F_2^{n}$ have degree exactly $d\ge2$, let $f=(-1)^{p}$, and let
$T=\Delta^d p$ be its top form. Then
\[
  \key{M(T)\;\le\;2^{\,d-1}\log_2 B(f)}.
\]
\end{lemma}

\begin{proof}
Fix $h_1,\dots,h_{d-2}\in\F_2^{n}$ and set
\[
  q\;=\;\Delta_{h_1}\cdots\Delta_{h_{d-2}}p,\qquad
  g\;=\;D_{h_1}\cdots D_{h_{d-2}}f\;=\;(-1)^{q},
\]
the second identity by \eqref{eq:iterated-deriv}. Each $\Delta$ lowers the degree by at
least one, so $\deg q\le d-(d-2)=2$.

We claim that the polar form of $q$ is exactly the contraction we are interested in:
\begin{equation}\label{eq:polar-is-contraction}
  \beta_q(u,v)\;=\;\Delta_u\Delta_vq\;=\;\Delta_u\Delta_v\Delta_{h_1}\cdots\Delta_{h_{d-2}}p
  \;=\;T(h_1,\dots,h_{d-2},u,v),
\end{equation}
where the middle equality is the definition of $q$ and the last follows from the
definition of $T$ and the commutativity of the difference operators.
If $\deg q<2$, then $\Delta_u\Delta_vq\equiv0$, and hence the contraction in \eqref{eq:polar-is-contraction} is zero as well.

We may therefore assume $\deg q=2$. Let $2r=\rank_{\F_2}\beta_q$, so $r\ge1$.
By Lemma~\ref{lem:quadratic} (resp.\ Lemma~\ref{lem:weight}(c))
\[
  B(g)\;=\;2^{r}.
\]
On the other hand \key{$g$ is a product of $2^{d-2}$ translates of $f$ by \eqref{eq:iterated-deriv}, so $B(g)\le B(f)^{2^{d-2}}$ by Fact~\ref{fact:product}
(resp.\ Lemma~\ref{lem:weight}(b))}. Combining all these, we have
\[
  \key{2^{r}\;\le\;B(f)^{2^{d-2}}
  \quad\Longrightarrow\quad
  r\;\le\;2^{\,d-2}\log_2B(f)
  \quad\Longrightarrow\quad
  \rank\beta_q=2r\;\le\;2^{\,d-1}\log_2 B(f).}
\]
Taking the maximum over $h_1,\dots,h_{d-2}$ proves the lemma. (The right-hand side is
positive by property (iv) of the convention above, so the inequality is not vacuous.)
\end{proof}

\begin{remark}\label{rem:no-compounding}
Lemma~\ref{lem:step1} is applied on every branch of the recursion. If $q$ is an affine
restriction of $p$ with $\deg q=j$, $2\le j\le d$, and $g=(-1)^{q}$, then
$M(\Delta^{j}q)\le2^{j-1}\log_2B(g)\le2^{j-1}\log_2B(f)$ by
Lemma~\ref{lem:restriction} (resp.\ Lemma~\ref{lem:weight}(a)). The logarithm is thus
always the \emph{original} one; no powers of $\log$ accumulate. This is the structural
difference from the proof of Theorem~\ref{thm:twxz}, whose recursion
re-introduces a logarithm at every level.
\end{remark}

\subsection{Step 2: generic rank versus binary evaluated rank}\label{sec:step2}

The algebraic input of Subsection~\ref{sec:step3} requires the rank of a contraction at a
\emph{generic} anchor tuple, whereas Lemma~\ref{lem:step1} bounds the rank at
\emph{binary} anchors. We now compare the two. Throughout this subsection $\F$ is an
arbitrary field.

\subsubsection{Multiplicity and the grid inequality}

\begin{definition}[Hasse derivatives and multiplicity]\label{def:hasse}
Let $P\in\F[x_1,\dots,x_m]$, and let $z=(z_1,\dots,z_m)$ be a separate tuple of
indeterminates, representing an increment. Write $\mathbf d=(d_1,\dots,d_m)\in\Z_{\ge0}^m$
for an exponent multi-index; boldface distinguishes it from the scalar degree $d$ used
elsewhere. \key{The \emph{Hasse derivatives} are defined by}
\[
  \key{P(x+z)=\sum_{\mathbf d\in\Z_{\ge0}^{m}}P^{(\mathbf d)}(x)z^{\mathbf d},
  \qquad z^{\mathbf d}=\prod_i z_i^{d_i}.}
\]
Thus $P^{(\mathbf d)}(x)$ is the coefficient of $z^{\mathbf d}$; \key{its \emph{order} is $|\mathbf d|=\sum_i d_i$}. \key{For a point $a=(a_1,\dots,a_m)\in\F^m$, the \emph{multiplicity}
$\mult_a P$ is the least total degree of a nonzero monomial of $P(a+z)$},
equivalently
\[
  \key{\mult_a P=\min\bigl\{|\mathbf d|:P^{(\mathbf d)}(a)\neq0\bigr\}}
\]
(with $\mult_a P=\infty$ if $P=0$). Here $x$ and $z$ are formal variables, while $a$
is a value assigned to $x$.
\end{definition}

\explain{In other words, Hasse derivatives are the coefficients of a \emph{formal Taylor expansion}, and multiplicity measures how many initial degree layers vanish at a point.
For nonzero $P$, multiplicity is zero exactly when $P(a)\neq0$; multiplicity $r$ means that all terms of total degree below $r$ vanish, but some degree-$r$ term survives.
In one variable, this is the usual root multiplicity: $P(x)=(x-a)^rQ(x)$ with $Q(a)\neq0$.}

\paragraph{A two-variable example.}
To better understand the above concepts and their properties, consider a monomial as an illustrative example. Let $P(x_1,x_2)=x_1^2x_2\in\F_2[x_1,x_2]$. With increments $(z_1,z_2)$,
\[
  P(x_1+z_1,x_2+z_2)=x_1^2x_2+x_1^2z_2+x_2z_1^2+z_1^2z_2.
\]
Then the nonzero Hasse derivatives are 
\[
P^{(0,0)}=x_1^2x_2,\quad P^{(0,1)}=x_1^2,\quad P^{(2,0)}=x_2,\quad \text{ and } P^{(2,1)}=1.
\] 
Reading the lowest total degree after translating to each point gives
\[
\begin{array}{c|c|c}
 a & P(a+(z_1,z_2)) & \mult_a P\\ \hline
 (0,0)&z_1^2z_2&3\\
 (1,0)&z_2+z_1^2z_2&1\\
 (0,1)&z_1^2+z_1^2z_2&2\\
 (1,1)&1+z_2+z_1^2+z_1^2z_2&0
\end{array}
\]
The three zero evaluations have different multiplicities: merely counting zeroes loses information about the order of vanishing. 

We next record two properties controlling the multiplicity $\mult_a P$, i.e.\ the
order of vanishing of $P$ at $a$, under Hasse differentiation and restriction to a line.

\begin{fact}\label{fact:hasse}
For any $P$, exponent multi-index $\mathbf d$ and point $a$:
\begin{enumerate}[label=(\alph*),itemsep=1pt]
  \item $\mult_a\bigl(P^{(\mathbf d)}\bigr)\ge\mult_a P-|\mathbf d|$;
  \item if $L$ is a line through $a$ and $P|_L$ denotes the restriction of $P$ to
        $L$ (a univariate polynomial), then $\mult_a(P|_L)\ge\mult_a P$
        whenever $P|_L\neq0$.
\end{enumerate}
\end{fact}

\begin{proof}
(a) Introduce another tuple $w=(w_1,\dots,w_m)$ of independent increment variables.
We expand the same polynomial $P(a+z+w)$ in two ways: first around $a+z$, and then
around $a$ with increment $z+w$. By the definition of Hasse derivatives,
\[
\begin{aligned}
 P(a+z+w)&=\sum_{\mathbf d''}P^{(\mathbf d'')}(a+z)w^{\mathbf d''},\\
 P(a+z+w)&=\sum_{\mathbf d''}P^{(\mathbf d'')}(a)(z+w)^{\mathbf d''}.
\end{aligned}
\]
In the first expansion, the coefficient of $w^{\mathbf d}$ is $P^{(\mathbf d)}(a+z)$.
In the second, the coordinatewise binomial theorem shows that the coefficient of
$w^{\mathbf d}$ in $(z+w)^{\mathbf d''}=\prod_i(z_i+w_i)^{d_i''}$ is
$\binom{\mathbf d''}{\mathbf d}z^{\mathbf d''-\mathbf d}$ when
$\mathbf d''\ge\mathbf d$ coordinatewise, and zero otherwise.
Here $\binom{\mathbf d''}{\mathbf d}:=\prod_i\binom{d_i''}{d_i}$ is the
multi-index binomial coefficient.
Comparing coefficients and putting $\mathbf d'=\mathbf d''-\mathbf d$ therefore gives
\[
 \key{P^{(\mathbf d)}(a+z)}
 =\sum_{\mathbf d''\ge\mathbf d}\binom{\mathbf d''}{\mathbf d}
   P^{(\mathbf d'')}(a)z^{\mathbf d''-\mathbf d} \
 \key{ =\sum_{\mathbf d'}\binom{\mathbf d'+\mathbf d}{\mathbf d} P^{(\mathbf d'+\mathbf d)}(a)z^{\mathbf d'}.}
\]
This argument uses only polynomial expansions and coefficient comparison, so it is valid
in every characteristic without division by factorials.

Now if $P^{(\mathbf d)}=0$, its multiplicity is infinite and (a) is immediate. Otherwise,
put $t=\mult_a(P^{(\mathbf d)})$ and \key{choose a nonzero term $z^{\mathbf d'}$ of lowest total degree} in $P^{(\mathbf d)}(a+z)$, so $|\mathbf d'|=t$. Its coefficient in the
formula above is nonzero, hence $P^{(\mathbf d'+\mathbf d)}(a)\neq0$. Thus \key{$P(a+z)$ has a nonzero term of degree $|\mathbf d'+\mathbf d|=t+|\mathbf d|$,} and thus
\[ 
\key{\mult_a P\le t + |\mathbf d| = \mult_a(P^{(\mathbf d)}) + |\mathbf d|.}
\]

(b) Assume $P|_L\neq0$ and put $r=\mult_a P$. \key{Parametrise the line through $a$ by $x=a+\lambda v$}, where $v\in\F^m$ is nonzero. By the definition of multiplicity,
\[
  P(a+z)=\sum_{|\mathbf d'|\ge r}c_{\mathbf d'}z^{\mathbf d'},
  \qquad c_{\mathbf d'}=P^{(\mathbf d')}(a).
\]
Substituting $z=\lambda v$ gives
\[
  \key{P(a+\lambda v)=\sum_{|\mathbf d'|\ge r}
  c_{\mathbf d'}v^{\mathbf d'}\lambda^{|\mathbf d'|}.}
\]
\key{Every term has $\lambda$-degree at least $r$. Terms may vanish or cancel, but no term of degree below $r$ can appear.} Thus the nonzero univariate restriction has multiplicity
at least $r$ at $\lambda=0$, i.e.\ $\mult_a(P|_L)\ge\mult_a P$.
\end{proof}


We will need to bound the sum of the multiplicities of a polynomial over a finite
Cartesian grid. To see what such a bound might look like on the binary grid,
consider the monomial $P(x)=\prod_{i=1}^n x_i^{e_i}$, where $e_i\ge0$ and $D=\sum_i e_i$ is its total degree. At a binary point $a\in\{0,1\}^n$, a factor $x_i^{e_i}$ contributes multiplicity $e_i$ when $a_i=0$ and zero when $a_i=1$. Multiplicities add under products, so
\[
  \mult_a P=\sum_{i:a_i=0}e_i,\qquad
  \Ex_{a\in\{0,1\}^n}[\mult_a P]=\frac12\sum_i e_i=\frac D2,
\]
since each coordinate of a uniformly random binary point is zero with probability $1/2$. Thus this monomial attains the bound $\sum_a\mult_a P=D\,2^{n-1}$.
For a general polynomial, cancellation between terms after translation can increase multiplicities at individual points. Nevertheless, the following lemma, due to Dvir, Kopparty, Saraf and Sudan~\cite{DKSS}, shows that the average multiplicity on the binary grid is at most $D/2$ for every nonzero polynomial of total degree at most $D$. More generally, it bounds the sum of the multiplicities over a finite Cartesian grid in terms of the total degree and the grid size, as shown in the next lemma.

\begin{lemma}[Multiplicity grid inequality~\cite{DKSS}]\label{lem:grid}
Let $P\in\F[x_1,\dots,x_n]$ be nonzero of total degree at most $D$ and let
$S\subseteq\F$ be finite. Then
\[
  \key{\sum_{a\in S^{n}}\mult_a P\;\le\;D\,|S|^{n-1}.}
\]
\end{lemma}

\subsubsection{The rank comparison}

Fix disjoint blocks of formal variables $\mathbf t_1,\dots,\mathbf t_k$, each of length
$n$, with $\mathbf t_i=(t_{i,1},\dots,t_{i,n})$, whose coordinates are algebraically
independent indeterminates, and put $K=\F(\mathbf t_1,\dots,\mathbf t_k)$. Polynomial
rings and function fields in these blocks are generated by all their coordinates.
The blocks $\mathbf t_i$ are formal, not vectors in $\F^{n}$; we use
$\mathbf a_i\in\F^{n}$ for values assigned to them.

\begin{definition}
A matrix $N=N(\mathbf t_1,\dots,\mathbf t_k)$ with entries in
$\F[\mathbf t_1,\dots,\mathbf t_k]$ is \emph{multi-affine} if every entry has total
degree at most one in each block. We call $N$ \emph{multilinear} if its entries are
linear and homogeneous in each block, as in our application to contractions of $T$.
Its \emph{generic rank} and maximal
\emph{binary evaluated rank} are, respectively,
\[
\begin{aligned}
  \key{R(N)=\rank_K N} \quad \text{ and }\quad 
  \key{M(N)=\max_{\mathbf a_i\in\{0,1\}^{n},\ 1\le i\le k}
  \rank_{\F} N(\mathbf a_1,\dots,\mathbf a_k).}
\end{aligned}
\]
\end{definition}

Here $\rank_K$ is the ordinary matrix rank over the rational function field $K$, with linear dependence computed using coefficients in $K$. Equivalently, $R(N)$ is the largest size of a minor whose determinant is a nonzero formal polynomial; that determinant may nevertheless vanish at every binary assignment.

For each binary assignment, substitute $\mathbf t_i=\mathbf a_i$ into $N$ and compute the ordinary matrix rank over $\F$. The word ``binary'' refers only to the assigned coordinates: the entries of the evaluated matrix lie in $\F$ and need not be $0$ or $1$, since the polynomial coefficients are arbitrary elements of $\F$. 
The entries of $N$ are polynomials, so these evaluations are always defined. The evaluated ranks are computed over $\F$, whereas the generic rank $R(N)$ is computed over $K$.

For multilinear matrices over $\F_2$, the following lemma is the second inequality
of Moshkovitz and Zhu~\cite[Theorem~1.1]{MoshkovitzZhu}, with $k$ variable blocks.
Their commutative rank $\mathrm{CR}(N)$ is our $R(N)$, and their max-rank
$\mathrm{MR}(N)$ is our $M(N)$, since binary assignments exhaust $\F_2^n$.

\begin{lemma}[Step 2]\label{lem:step2}
Let $\F=\F_2$ and let $N$ be a multilinear matrix in $k$ blocks. Then
\key{$R(N)\le2^{k}M(N)$}.
\end{lemma}

For completeness, we provide a proof of this special case in
Appendix~\ref{app:step2-proof}.

To see why a loss is necessary, consider one block $\mathbf t=(t_1,\dots,t_n)$ and
let $N_0(\mathbf t)$ be the diagonal matrix whose entries are all the $2^n-1$ nonzero
linear forms $\ell_v(\mathbf t)=\sum_i v_it_i$, indexed by $v\in\F_2^n\setminus\{0\}$.
For $n=2$, this is simply $\operatorname{diag}(t_1,t_2,t_1+t_2)$.
Each diagonal entry is a nonzero polynomial, so $R(N_0)=2^n-1$.
On the other hand, at a nonzero binary point $a$, exactly $2^{n-1}$ of the forms
satisfy $\ell_v(a)=1$, while $N_0(0)=0$. Thus
\[
  M(N_0)=2^{n-1},\qquad
  \frac{R(N_0)}{M(N_0)}=2-2^{1-n}\longrightarrow2.
\]
Moreover, each nonzero linear form is one at exactly half of the binary points,
so the average evaluated rank is exactly $R(N_0)/2$.
This family is the binary case of the example in
Moshkovitz and Zhu~\cite[Proposition~5.5]{MoshkovitzZhu}.
For general multilinear matrices, their Corollary~5.3 gives
\[
  M(N)\ge
  \Ex_{\mathbf a_1,\dots,\mathbf a_k\in\F_2^n}
  [\rank_{\F_2}N(\mathbf a_1,\dots,\mathbf a_k)]
  \ge2^{-k}R(N).
\]
In other words, each additional binary block costs at most a factor of two in
the average rank, explaining the factor $2^k$ in Lemma~\ref{lem:step2}.

\begin{corollary}\label{cor:step1-2}
Let $p$ on $\F_2^{n}$ have degree exactly $d\ge3$, let $T=\Delta^{d}p$ and
$f=(-1)^{p}$, and let $R$ denote the generic
rank of $T(\mathbf t_1,\dots,\mathbf t_{d-2},-,-)$ over
$K=\F_2(\mathbf t_1,\dots,\mathbf t_{d-2})$, where the $\mathbf t_i$ are independent
formal anchor blocks of length $n$. Then
\[
  R\;\le\;2^{\,d-2}M(T)\;\le\;2^{\,2d-3}\log_2 B(f).
\]
\end{corollary}

\begin{proof}
Let $N(\mathbf t_1,\dots,\mathbf t_{d-2})$ be the matrix of the bilinear form
$T(\mathbf t_1,\dots,\mathbf t_{d-2},-,-)$. By Lemma~\ref{lem:Tj}, $T$ is multilinear,
so the entries of $N$ are linear and homogeneous in each of the $k=d-2$ formal blocks.
Evaluating $\mathbf t_i=h_i$,
with $h_i\in\F_2^n$, gives the matrix of $T(h_1,\dots,h_{d-2},-,-)$; hence
$R=\rank_KN$ and $M(N)=M(T)$. Apply Lemma~\ref{lem:step2} and then Lemma~\ref{lem:step1}.
\end{proof}

\subsection{Step 3: generic radicals satisfy a strong vanishing identity}\label{sec:step3}

This subsection is the algebraic heart of the proof. Let $\F$ be a field, $V=\F^{n}$, and let
$T\colon V^{d}\to\F$ be a symmetric $d$-linear form, $d\ge3$. We extend $T$ to any field
extension by the same coefficients. Only the symmetry of $T$ is used here; the alternating
property plays no role in this subsection.

The rank and radical of a bilinear form are independent of the choice of basis.
In any basis, the rank of the form equals that of its Gram matrix, and the coordinate
map identifies its radical with the kernel of that matrix. We therefore use the
bilinear form itself, without introducing separate notation for its matrix.

\begin{theorem}[Step 3]\label{thm:step3}
Let $\mathbf t_i=(t_{i,1},\dots,t_{i,n})$, $1\le i\le d-2$, be formal anchor blocks
whose coordinates are all independent indeterminates. Put
$K=\F(\mathbf t_1,\dots,\mathbf t_{d-2})$ and let
\[
  \key{A\;=\;T(\mathbf t_1,\dots,\mathbf t_{d-2},-,-)\colon K^{n}\times K^{n}\to K.}
\]
Put
\[
  \key{R=\rank_K A,\qquad U=\ker A\le K^{n}.}
\]
Then \key{$\dim_K U=n-R$} and
\[
  \key{T\bigl(K^{n},\,U,\,U,\dots,U\bigr)\;=\;0} ,
\]
i.e.\ $T(v,u_1,\dots,u_{d-1})=0$ for every $v\in K^{n}$ and $u_1,\dots,u_{d-1}\in U$.
\end{theorem}

Note that the conclusion is strictly stronger than $T(U^{d})=0$: one of the $d$ slots is
allowed to range over the whole space. Only the weaker conclusion is used in
Subsection~\ref{sec:proof}, but the stronger one is what the induction naturally produces.

\begin{proof}
If $R=0$ then $A=0$; since $\mathbf t_1,\dots,\mathbf t_{d-2}$ are independent indeterminates and $T$ is
multilinear, this forces $T=0$ and the conclusion is trivial. Assume $R\ge1$.

\key{The first assertion is the rank--nullity theorem} over $K$. For the second, fix a complement $S$ of
$U$ in $K^{n}$, so $\dim_K S=R$ and \key{$K^{n}=S\oplus U$}. When using block matrices,
we identify bilinear forms with their Gram matrices in a basis adapted to this
decomposition. In particular,
\[
  \key{A\;=\;\begin{pmatrix}P_0 & 0\\ 0& 0\end{pmatrix},\qquad P_0\in K^{R\times R}\ \text{invertible}},
\]
because $U$ is exactly the radical of $A$ (so the off-diagonal blocks and the lower-right
block vanish) and the restriction of $A$ to $S$ is non-degenerate.

\paragraph{Perturbation.}
Fix direction vectors $w_1,\dots,w_{d-2}\in K^{n}$.
Let $s_1,\dots,s_{d-2}$ be new independent indeterminates, distinct from the
coordinates of the formal anchor blocks $\mathbf t_i$, and consider
\[
  \key{A(s)\;=\;T\bigl(\mathbf t_1+s_1w_1,\;\dots,\;
  \mathbf t_{d-2}+s_{d-2}w_{d-2},\;-,\;-\bigr),}
\]
a bilinear form over $K(s)=K(s_1,\dots,s_{d-2})=\F(\mathbf t_1, ..., \mathbf t_{d-2},s)$. By multilinearity of $T$, every entry of
$A(s)$ is affine-linear in each $s_i$; we may therefore expand
\begin{equation}\label{eq:expansion}
  A(s)\;=\;\sum_{I\subseteq\{1,\dots,d-2\}}A_I\prod_{i\in I}s_i ,
\end{equation}
a \emph{finite} expansion in which the coefficient $A_I$ is obtained by substituting $w_i$
into slot $i$ for $i\in I$ and $\mathbf t_i$ for $i\notin I$. Explicitly,
\begin{equation}\label{eq:coeff-meaning}
  \key{A_I\;=\;T\bigl(\,b_1,\dots,b_{d-2},\,-,\,-\,\bigr),
  \qquad b_i=\begin{cases}w_i,& i\in I,\\ \mathbf t_i,& i\notin I.\end{cases}}
\end{equation}
In particular $A_\emptyset=A$.

\paragraph{The rank does not grow.}
Since the coordinates of $\mathbf t_1,\dots,\mathbf t_{d-2}$ are independent
indeterminates and $\rank_K A=R$, \key{every $(R+1)\times(R+1)$ minor of the Gram matrix
of $A$ in the standard basis vanishes identically as a polynomial} in the anchor
coordinates. (A nonzero such minor would be nonzero at the generic point,
contradicting $\rank_K A=R$.) Substituting $\mathbf t_i\mapsto\mathbf t_i+s_iw_i$ in these
polynomial identities therefore gives \key{$\rank_{K(s)}A(s)\le R$}.

\paragraph{Schur complement.}
Write $A(s)$ in the block form induced by $K^{n}=S\oplus U$:
\[
  \key{A(s)\;=\;\begin{pmatrix}P(s)&B(s)\\ B(s)^{\mathsf T}&E(s)\end{pmatrix},
  \qquad P(0)=P_0,\quad B(0)=E(0)=0 .}
\]
The lower-left block is $B(s)^{\mathsf T}$ because symmetry of $T$ in its two free slots makes $A(s)$ symmetric. 
The values at $s=0$ are read off from $A(0)=A_\emptyset=A$ 
(the adapted basis does not depend on $s$). 
Write $B_I$ and $E_I$ for the corresponding blocks of the coefficient matrix $A_I$; its lower-left block is likewise $B_I^{\mathsf T}$.
Since $P(0)=P_0$ is invertible, $\det P(s)$ is a nonzero element of $K[s]$, so $P(s)$ is invertible over $K(s)$. 
The block-rank identity gives
\[
  \key{\rank_{K(s)}A(s)=R+\rank_{K(s)}\bigl(E(s)-B(s)^{\mathsf T}P(s)^{-1}B(s)\bigr).}
\]
\key{As $\rank_{K(s)}A(s)\le R$, the Schur complement must vanish}:
\begin{equation}\label{eq:schur}
  E(s)\;=\;B(s)^{\mathsf T}\,P(s)^{-1}\,B(s)\qquad\text{identically in }K(s).
\end{equation}
Because $P(s)$ is invertible at $s=0$, all entries of $P(s)^{-1}$ are regular at the origin.
Put $J(s)=P_0^{-1}(P_0-P(s))$, so $P(s)=P_0(I-J(s))$ and
$P(s)^{-1}=(I-J(s))^{-1}P_0^{-1}$. The matrix geometric series then gives
\begin{equation}\label{eq:neumann}
  \key{P(s)^{-1}\;=\;\sum_{\nu\ge0}\bigl(P_0^{-1}(P_0-P(s))\bigr)^{\nu}P_0^{-1}
  \qquad\text{in } K[[s_1,\dots,s_{d-2}]],}
\end{equation}
Since $J(0)=0$, every entry of $J(s)^\nu$ has no terms of total degree below $\nu$.
Thus a coefficient of total degree $D$ receives contributions only from $\nu\le D$,
so its computation involves only finitely many terms. This makes the expansion
formally well defined in $K[[s]]$, without any analytic convergence requirement.
We may therefore compare coefficients of monomials on the two sides of
\eqref{eq:schur} inside the formal power series ring $K[[s]]$, where such an expansion is unique.
By \eqref{eq:expansion}, \key{each entry of the left-hand side $E(s)$ is multilinear in
$s_1,\dots,s_{d-2}$}, i.e.\ has degree at most one in each $s_i$.

\paragraph{Simultaneous induction.}
We prove by induction on $|I|$ the following two assertions, for all choices of the
perturbation vectors $w_1,\dots,w_{d-2}$:
\begin{enumerate}[label=(\arabic*),itemsep=2pt]
  \item\label{it:E} \key{if at most one of the vectors $\{w_i\}_{i\in I}$ lies outside $U$,
        then $E_I=0$;}
  \item\label{it:BC} \key{if all of the vectors $\{w_i\}_{i\in I}$ lie in $U$, then $B_I=0$ and $B_I^{\mathsf T}=0$.}
\end{enumerate}

\emph{Base case $|I|=1$, say $I=\{i\}$.} Since $B(s)$ has zero constant matrix, all entries
of both $B(s)$ and $B(s)^{\mathsf T}$ lie in the ideal $(s_1,\dots,s_{d-2})$ of $K[[s]]$.
Hence every entry of $B(s)^{\mathsf T}P(s)^{-1}B(s)$ lies in the square of that ideal and
has no monomial of \key{total degree one}. By \eqref{eq:schur}, $E_{\{i\}}=0$. This is
\ref{it:E} (indeed unconditionally at $|I|=1$).

\key{Assertion \emph{\ref{it:E} at size $|I|$ implies assertion \ref{it:BC} at size $|I|$.}}
Suppose all $w_i\in U$ for $i\in I$, and write $I=\{i_1,\dots,i_\ell\}$.
For notational convenience, after relabelling the anchor blocks and perturbation
vectors, assume without loss of generality that \key{$I=\{1,2,\dots,\ell\}$}.
Entries of $B_I$ are, by \eqref{eq:coeff-meaning}, the values
\[
  \key{T\bigl(w_1,\dots,w_\ell,\mathbf t_{\ell+1},\dots,\mathbf t_{d-2},\sigma,u\bigr),
  \qquad \sigma\in S,\ u\in U,}
\]
with empty index ranges omitted. \key{By symmetry} we may interchange $\sigma$ with $w_1$,
obtaining
\[
  \key{T\bigl(\sigma,w_2,\dots,w_\ell,\mathbf t_{\ell+1},\dots,\mathbf t_{d-2},w_1,u\bigr).}
\]
Set $w'_1=\sigma$ and $w'_i=w_i$ for $2\le i\le\ell$. Among the indices in $I$,
\key{only $w'_1$ may lie outside $U$}, so assertion \ref{it:E} at size $\ell$ gives
$E_I=0$ for these modified directions. \key{Since $w_1,u\in U$}, the displayed value
is an evaluation of this \key{$U\times U$ block} by \eqref{eq:coeff-meaning}, and \key{hence is zero}.
By symmetry, the original value is also zero.
Thus every $S\times U$-value of the original coefficient form vanishes, so
$B_I=B_I^{\mathsf T}=0$, proving assertion \ref{it:BC}.

\key{\emph{Induction step for assertion \ref{it:E}.}}
Let $|I|=\ell\ge2$ and suppose at most one $w_i$, $i\in I$, lies outside $U$. \key{Expand the
right-hand side of \eqref{eq:schur} using \eqref{eq:neumann}.} Because $B(s)$ and its
transpose have zero constant term, \key{any contribution} to the coefficient of the squarefree
monomial $\prod_{i\in I}s_i$ \key{must use a term $B_J^{\mathsf T}$ with
$\emptyset\neq J\subseteq I$, a term $B_K$ with $\emptyset\neq K\subseteq I$, and a term of
\eqref{eq:neumann} indexed by the remaining variables}, with $J$, $K$ and the index
multiset coming from $P(s)^{-1}$ pairwise disjoint and covering $I$. (Terms in which some
$s_i$ appears with exponent $\ge2$ cannot contribute to a squarefree monomial, so no case
is omitted.) Since $J$ and $K$ are nonempty and disjoint, both are \emph{proper} subsets
of $I$, so \key{$|J|,|K|<\ell$}.

By hypothesis at most one $w_i$ with $i\in I$ lies outside $U$. \key{As $J\cap K=\emptyset$, at
least one of $J$, $K$ consists entirely of indices $i$ with $w_i\in U$. If this holds for
$J$ then $B_J=0$, hence $B_J^{\mathsf T}=0$}, by \ref{it:BC} at size $|J|<\ell$; if it holds
for $K$ then $B_K=0$ and $B_K^{\mathsf T}=0$ by the same assertion. Either way \key{every contribution vanishes, so
$E_I=0$}.

\paragraph{Conclusion.}
For $d=3$, the desired identity $T(v,u,u')=0$ follows directly from the base case
of assertion \ref{it:E}. In general, take \key{$I=\{1,\dots,d-2\}$ and apply assertion
\ref{it:E}}. By \eqref{eq:coeff-meaning}, the $(u,u')$-entry of $E_I$ equals
$T(w_1,\dots,w_{d-2},u,u')$ for $u,u'\in U$. Choosing
$w_1=v\in K^{n}$ arbitrary and $w_2,\dots,w_{d-2}\in U$ -- so that at most one
perturbation vector lies outside $U$ -- we get
\[
  \key{T(v,w_2,\dots,w_{d-2},u,u')=0\qquad\text{for all }v\in K^{n},\ w_2,\dots,w_{d-2},u,u'\in U.}
\]
This is precisely $T(K^{n},U,\dots,U)=0$.
\end{proof}

\begin{remark}
Three features of the proof make it work in characteristic two. (i) No polarisation
identity of the form $T(x,\dots,x)=d!\cdot(\dots)$ is used, so no division by $d!$ occurs.
(ii) The vanishing of all $(R+1)$-minors is a statement about polynomials, established
before any substitution; it does not require the perturbed anchors to be generic. (iii)
The induction is on the size of the index set $I$, and assertions \ref{it:E} and
\ref{it:BC} are interleaved in a well-founded way (\ref{it:E} at size $\ell$ uses
\ref{it:BC} at sizes $<\ell$, and \ref{it:BC} at size $\ell$ uses \ref{it:E} at size
$\ell$, which was established first), so there is no circularity.
\end{remark}

\subsection{Step 4: descent to \texorpdfstring{$\F_2$}{F\_2} without losing dimension}\label{sec:step4}

Theorem~\ref{thm:step3} produces a subspace of $K^{n}$ with $K$ a rational function field.
For the tree construction we need a subspace of $\F_2^{n}$ of the same dimension over
its own coefficient field. In the following lemma the scalar indeterminates
$z_1,\dots,z_L$ can be the coordinates of the formal anchor blocks $\mathbf t_i$.

\begin{lemma}[Step 4]\label{lem:step4}
Let $T\colon(\F_2^{n})^{d}\to\F_2$ be multilinear with coefficients in $\F_2$, extended to
any field extension by the same coefficients. Let $K=\F_2(z_1,\dots,z_L)$ be a purely
transcendental extension and suppose $U\le K^{n}$ is a $K$-subspace with $\dim_K U=k$
and $T(K^{n},U,\dots,U)=0$.
\key{Then there is an $\F_2$-subspace $U'\le\F_2^{n}$ with $\dim_{\F_2}U'=\dim_K U=k$ and $T(\F_2^{n},U',\dots,U')=0$; in particular $T(U',\dots,U')=0$.}
\end{lemma}

\begin{proof}
\key{We eliminate the indeterminates $z_1,\dots,z_L$ one at a time}, preserving the subspace dimension over the current coefficient field and the vanishing identity.
\key{After all indeterminates have been eliminated, the coefficient field is $\F_2$, as desired.}
Put $z=z_1$ and $F=\F_2(z_2,\dots,z_L)$, so
$K=F(z)$; when $L=1$, this means $F=\F_2$. We first eliminate $z_1$ by constructing
an $F$-subspace $U_0\le F^n$ with $\dim_F U_0=k$ and $T(F^n,U_0,\dots,U_0)=0$.
The argument below works for any coefficient field $F$ containing $\F_2$, so the same
step can then be repeated for the remaining indeterminates.

Choose a basis matrix $\Lambda\in F(z)^{n\times k}$ of $U$, with $\rank_{F(z)}\Lambda=k$.
Write $\Lambda=(\Lambda_1,\dots,\Lambda_k)$, where $\Lambda_c$ denotes the
$c$-th column of $\Lambda$. Then
\begin{equation}\label{eq:vanish-cols}
  T(e_{i_0},\Lambda_{c_1},\dots,\Lambda_{c_{d-1}})=0
  \quad\text{for all }i_0\in[n],\ c_1,\dots,c_{d-1}\in[k] ,
\end{equation}
which by multilinearity is equivalent to $T(K^{n},U,\dots,U)=0$.

Let $v$ denote the $z$-adic valuation on $F(z)$
, and let $\mathcal O=\{\lambda\in F(z):v(\lambda)\ge0\}$ be its valuation ring, with residue map
$\lambda\mapsto\lambda|_{z=0}$ onto $F$. \key{Among the $\binom{n}{k}$ maximal ($k\times k$) minors of $\Lambda$}, at least one is nonzero; \key{choose a row set $\rho\subseteq[n]$, $|\rho|=k$, whose minor $\det\Lambda_\rho$ is nonzero and has
\emph{minimal} valuation $v(\det\Lambda_\rho)$ among all nonzero maximal minors.} Replace $\Lambda$ by
\[
  \Lambda^{\ast}\;=\;\Lambda\cdot\bigl(\Lambda_{\rho}\bigr)^{-1} .
\]
This is a right multiplication by an invertible $k\times k$ matrix, i.e.\ a change of
basis of the column space, so the subspace $U$ and hence the validity of \eqref{eq:vanish-cols} are unchanged. 
The rows of $\Lambda^{\ast}$ indexed by $\rho$ form the
identity matrix. By Cramer's rule, the $(i,c)$ entry of
$\Lambda\cdot\mathrm{adj}(\Lambda_\rho)$ is the determinant of $\Lambda_\rho$ with its
$c$-th row replaced by the $i$-th row of $\Lambda$, i.e.\ $\pm\det\Lambda_{\rho'}$ for the
row set $\rho'$ so obtained; hence each entry of $\Lambda^{\ast}$ equals
$\pm\det\Lambda_{\rho'}/\det\Lambda_{\rho}$. 
Since $\rho$ was chosen to minimise the valuation among all nonzero maximal minors,
$v(\det\Lambda_{\rho'})\ge v(\det\Lambda_\rho)$ whenever $\det\Lambda_{\rho'}\ne0$.
The valuation of a quotient is the difference of the valuations, so
\[
  \key{v\!\left(\frac{\det\Lambda_{\rho'}}{\det\Lambda_\rho}\right)
  =v(\det\Lambda_{\rho'})-v(\det\Lambda_\rho)\ge0.}
\]
If the numerator is zero, the corresponding entry is zero as well.
Thus \key{every entry of $\Lambda^{\ast}$ lies in $\mathcal O$ (i.e. has nonnegative valuation) and can be evaluated at $z=0$.}

\key{Now eliminate $z$ by evaluating the entries of $\Lambda^{\ast}$ at $z=0$}, which is
well defined since they all lie in $\mathcal O$. Let
$\Lambda^{0}=\Lambda^{\ast}|_{z=0}\in F^{n\times k}$.
\key{Its rows indexed by $\rho$ are still the identity,} so $\rank_F\Lambda^{0}=k$ and its columns span an $F$-subspace $U_0$ with $\dim_FU_0=k$. 
It remains to check that the vanishing identity is preserved. Each expression
$T(e_{i_0},\Lambda^{\ast}_{c_1},\dots,\Lambda^{\ast}_{c_{d-1}})$ is a finite sum of
products of entries of $\Lambda^{\ast}$ and coefficients of $T$. All these factors
lie in $\mathcal O$, so the expression itself lies in $\mathcal O$.
The residue map $\phi\colon\mathcal O\to F$, given by evaluation at $z=0$, preserves
addition and multiplication and fixes the coefficients of $T$, which lie in $\F_2$.
Applying it to \eqref{eq:vanish-cols}, written for $\Lambda^{\ast}$, therefore gives
\[
\begin{aligned}
  0&=\phi\!\left(T(e_{i_0},\Lambda^{\ast}_{c_1},\dots,\Lambda^{\ast}_{c_{d-1}})\right)=T(e_{i_0},\Lambda^{0}_{c_1},\dots,\Lambda^{0}_{c_{d-1}})
\end{aligned}
\]
for all $i_0\in[n]$ and $c_1,\dots,c_{d-1}\in[k]$.
By multilinearity, this extends from the standard basis vectors and the columns
of $\Lambda^0$ to all vectors in $F^n$ and $U_0$, respectively.
Hence \key{$T(F^{n},U_0,\dots,U_0)=0$, with $\dim_FU_0=k$ as established above.}

The resulting $U_0$ is defined over $F=\F_2(z_2,\dots,z_L)$. If $L\ge2$, apply the
same construction to $U_0$, now with $z_2$ as the indeterminate and
$\F_2(z_3,\dots,z_L)$ as the coefficient field (equal to $\F_2$ when $L=2$).
Continue in this way, normalising the basis anew before each evaluation. Each step
preserves dimension over the current coefficient field and the vanishing identity,
since $T$ always has coefficients in $\F_2$. After eliminating all $L$ indeterminates,
the coefficient field is $\F_2$ and the resulting subspace $U'\le\F_2^n$ satisfies
$\dim_{\F_2}U'=\dim_K U=k$ and $T(\F_2^n,U',\dots,U')=0$.
\end{proof}

\paragraph{Remarks.}
\begin{enumerate}[label=(\roman*),itemsep=4pt]
  \item \emph{Successive descent.} Each step constructs a new subspace over a smaller
    coefficient field, following the chain
    \[
      \F_2(z_1,\dots,z_L)\longrightarrow\F_2(z_2,\dots,z_L)
      \longrightarrow\cdots\longrightarrow\F_2(z_L)\longrightarrow\F_2.
    \]
    Each new subspace has dimension $k$ over its current coefficient field and
    satisfies the same vanishing condition for $T$.
  \item \emph{The core of the argument.} Normalisation makes evaluation well defined;
    the identity submatrix prevents any loss of dimension; and evaluation preserves
    addition and multiplication, hence the vanishing identity.
  \item \emph{Why a fraction field?} Normalisation requires inverting $\Lambda_\rho$,
    which involves division by $\det\Lambda_\rho$. Thus even if $\Lambda$ initially
    has polynomial entries, $\Lambda^\ast$ may have rational entries. This is why we
    work over a rational function field rather than a polynomial ring. For evaluation,
    it suffices that the normalised entries have no pole at the chosen point;
    they need not be polynomials.
\end{enumerate}

\paragraph{ }

\begin{corollary}\label{cor:c-le-R}
Let $p$ on $\F_2^{n}$ have degree exactly $d\ge3$, let $T=\Delta^{d}p$ and
$f=(-1)^{p}$. Let $\mathbf t_1,\dots,\mathbf t_{d-2}$ be independent formal anchor
blocks of length $n$, put $K=\F_2(\mathbf t_1,\dots,\mathbf t_{d-2})$, and let $N$
be the matrix of $T(\mathbf t_1,\dots,\mathbf t_{d-2},-,-)$ in the standard basis of $K^n$.
With $R=\rank_KN$, we have
\[
  c(T)\;\le\;R\;\le\;2^{\,2d-3}\log_2 B(f).
\]
\end{corollary}

\begin{proof}
We first construct the vanishing subspace, using Steps~3 and~4, and then bound its
codimension using Steps~2 and~1.

\emph{Step 3.} Apply Theorem~\ref{thm:step3} with $\F=\F_2$ to the $d$-linear form $T$.
The generic two-slot contraction $A=T(\mathbf t_1,\dots,\mathbf t_{d-2},-,-)$ has
matrix $N$, hence $\rank_K A=R$. The theorem gives $U=\ker A\le K^n$ satisfying
\[
  \dim_K U=n-R,\qquad T(K^n,U,\dots,U)=0.
\]

\emph{Step 4.} Lemma~\ref{lem:step4}, applied to the coordinates of the formal anchor
blocks, gives $U'\le\F_2^n$ with
\[
  \dim_{\F_2}U'=\dim_K U=n-R,\qquad T(\F_2^n,U',\dots,U')=0.
\]
In particular $T(U',\dots,U')=0$, so the definition of $c(T)$ yields
\[
  c(T)\le n-\dim_{\F_2}U'=R.
\]

\emph{Step 2.} The entries of $N$ are linear and homogeneous in each of the $k=d-2$
formal blocks. Evaluating $\mathbf t_i=h_i\in\F_2^n$ gives the matrix of the binary
contraction $T(h_1,\dots,h_{d-2},-,-)$; hence $M(N)=M(T)$.
Lemma~\ref{lem:step2} therefore gives
\[
  R=\rank_KN\le2^{d-2}M(T).
\]

\emph{Step 1.} By Lemma~\ref{lem:step1}, $M(T)\le2^{d-1}\log_2B(f)$.
Combining the four steps, we obtain
\[
  \key{c(T)\le n-\dim_{\F_2}U' = n-\dim_{K}U =R
  \le2^{d-2}M(T)\le2^{2d-3}\log_2B(f).}
\]
By Lemma~\ref{lem:degree-drop}, querying a basis of $U'^\perp$ uses exactly $R$
parity queries and lowers the degree on every branch from $d$ to at most $d-1$.
\end{proof}

\subsection{Proof of the main theorem}\label{sec:proof}
Now we are ready to prove the main theorem.
\begin{proof}[Proof of Theorems~\ref{thm:main} and~\ref{thm:main-weight}]
Let $p$ have degree exactly $d\ge2$, let $f=(-1)^{p}$, and let
$B=B(f)\in\{\|\widehat f\|_1,W(f)\}$. We apply Lemma~\ref{lem:recursion} with
\[
  \Gamma(2)\;=\;\log_2 B,\qquad
  \Gamma(j)\;=\;2^{\,2j-3}\log_2 B\quad(j\ge3).
\]

\emph{Verification for $3\le j\le d$.} Let $q$ be an affine restriction of $p$ of degree exactly
$j$, on an affine subspace of dimension $m$, and let $g=(-1)^{q}$. By
Lemma~\ref{lem:restriction} (resp.\ Lemma~\ref{lem:weight}(a)) we have
$B(g)\le B(f)$. Apply Corollary~\ref{cor:c-le-R} to $q$, with the corollary's ambient
dimension and degree parameter set to $m$ and $j$, respectively. This gives
\[
  c(\Delta^{j}q)\;\le\;2^{\,2j-3}\log_2B(g)\;\le\;2^{\,2j-3}\log_2 B\;=\;\Gamma(j).
\]

\emph{Verification for $j=2$.} Let $q$ be a degree-$2$ affine restriction of $p$, with polar
rank $2r$, and let $g=(-1)^q$. By Lemma~\ref{lem:quadratic} (resp.\ Lemma~\ref{lem:weight}(c))
we have $2^{r}=B(g)\le B$, i.e.\ $r\le\log_2B$. By Proposition~\ref{prop:quadratic-exact},
$\Dxor(g)\le r+1\le\Gamma(2)+1$.

Lemma~\ref{lem:recursion} now gives
\[
\begin{aligned}
  \Dxor(f)&\;\le\;1+\sum_{j=2}^{d}\Gamma(j)
  \;=\;1+\Bigl(1+\sum_{j=3}^{d}2^{\,2j-3}\Bigr)\log_2 B \;=\;1+\frac{2^{\,2d-1}-5}{3}\log_2 B .
\end{aligned}
\]
Here the sum from $j=3$ to $d$ is empty when $d=2$.
This proves Theorem~\ref{thm:main-weight} with $B=W(f)$ and the first inequality of
Theorem~\ref{thm:main} with $B=\|\widehat f\|_1$. The second inequality of
Theorem~\ref{thm:main} follows from $\|\widehat f\|_1\le\sqrt{s}$, i.e.\
$\log_2\|\widehat f\|_1\le\half\log_2 s$, by \eqref{eq:parseval}.
\end{proof}

\begin{remark}
For $d=2$ the formula gives $(2^{3}-5)/6=\half$, i.e.\
$\Dxor(f)\le1+\half\log_2 s$, which is exactly optimal by
Proposition~\ref{prop:quadratic-exact}. For $d=3$ it gives $\Dxor(f)\le1+4.5\log_2 s$.
\end{remark}

\begin{remark}[The construction is not algorithmic]
The proof establishes the \emph{existence} of a shallow parity decision tree. Steps 3 and 4
involve ranks and radicals over a rational function field in $\Theta(nd)$ indeterminates,
and we make no claim about the time required to find the tree from a succinct description
of $p$.
\end{remark}

\subsubsection{Range of applicability}

By Lemma~\ref{lem:d-le-logs} we always have $4^{d}\le s^{2}$, so
Theorem~\ref{thm:main} is informative only when $d$ is substantially smaller than
$\log s$; once $4^{d}\log_2 s\gtrsim\sqrt s$, i.e.\ roughly $d\ge\tfrac14\log_2 s$,
the unconditional bound $\Dxor=O(\sqrt s)$ of~\cite{TWXZ13} is stronger. The two
bounds leave open the intermediate range
\[
  \omega(\log\log s)\;\le\;d\;\le\;O(\log s),
\]
and we are not aware of any bound that beats both endpoints inside it.

\section{Concluding remarks}\label{sec:conclusion}

For a Boolean phase $f=(-1)^p$ with $\deg p=d\ge2$, we have proved
\[
  \Dxor(f)\le1+\frac{2^{2d-1}-5}{3}\log_2 B(f),
  \qquad B(f)\in\{\|\widehat f\|_1,W(f)\}.
\]
In particular, $\Dxor(f)=O(4^d\log_2 s(f))$, with the same asymptotic upper bound
for deterministic communication complexity of the associated XOR function.
The bound is optimal for quadratic phases and establishes the XOR log-rank
conjecture throughout the degree range $d=O(\log\log s(f))$.

The proof combines four ingredients: control of binary contraction ranks by the
spectral norm or threshold weight, a comparison between generic and binary ranks,
a strong vanishing identity for the radical of a generic contraction, and descent
to $\F_2$ preserving dimension. Together they produce a binary subspace on which
the top form vanishes, allowing a fixed batch of parity queries to lower the
degree on every branch. Monotonicity of $B(f)$ under affine restriction then
lets us sum the costs over degree levels without accumulating powers of the
logarithm.

Several problems remain open. First, our result establishes the existence of a shallow
parity decision tree but leaves its efficient construction open. Second, can the
classical bound $O(4^d\log_2 s(f))$ be improved to $O(2^d\log_2 s(f))$, or better?
For comparison, Zhang~\cite{Zhang14} proved that the exact quantum communication
complexity of $F(x,y)=f(x\oplus y)$ satisfies $Q_E(F)=O(2^d\log_2 s(f))$.
Finally, could the log-rank conjecture fail in general, perhaps already for XOR functions?

\clearpage
\bibliographystyle{alpha}
\bibliography{references}

\clearpage
\appendix
\section{Proof of the binary rank comparison}\label{app:step2-proof}

For completeness, we give a proof of Lemma~\ref{lem:step2} using the multiplicity
grid inequality of Dvir, Kopparty, Saraf and Sudan~\cite{DKSS}.
The argument in fact proves $R(N)\le2^kM(N)$ for multi-affine matrices over any
field $\F$, with $M(N)$ defined using binary assignments as in
Subsection~\ref{sec:step2}. Allowing an arbitrary coefficient field is useful
because the induction treats the remaining formal blocks as part of that field.

\begin{proof}
We prove the statement by induction on $k$. For $k=0$, both ranks are the ordinary
matrix rank over $\F$, so the assertion is immediate. For $k=1$, work over an
arbitrary base field $\F$ and write $t=(t_1,\dots,t_n)$
for the single formal block and $a=(a_1,\dots,a_n)\in\{0,1\}^{n}$ for a binary
assignment. Put $R=R(N)$. If $R=0$, there is nothing to prove, so assume $R\ge1$.
Choose an $R\times R$ submatrix whose determinant $P\in\F[t]$ is not identically zero.
Since each entry of $N$ is affine-linear in $t$, we have $\deg P\le R$.
Every $(R+1)$-minor of $N$ is identically zero, since its nonvanishing would contradict
$\rank_{\F(t)}N=R$. Thus $r_a:=\rank_{\F}N(a)\le R$ for every $a$.

\emph{Claim: for every $a\in\{0,1\}^{n}$, $\mult_a P\ge R-r_a$.}
Let $N'$ be the chosen $R\times R$ submatrix. Then $\rank_{\F}N'(a)\le r_a$,
so there is an invertible matrix $C$ over $\F$ such that at least
$R-r_a$ rows of $CN'(a)$ vanish. Replacing $N'$ by $CN'$ multiplies $P$ by the
nonzero constant $\det C$ and changes neither its degree nor its multiplicities.
Writing $t=a+z$, each entry in those rows of $CN'(a+z)$ has zero constant term.
Every monomial in the Leibniz expansion of $\det(CN'(a+z))$ picks one entry from
each row, hence contains at least $R-r_a$ factors with zero constant term.
Thus every nonzero monomial of $P(a+z)$ has total degree at least $R-r_a$,
proving the claim.

Applying Lemma~\ref{lem:grid} with $S=\{0,1\}$ and $D=R$ gives
\[
  \sum_{a\in\{0,1\}^{n}}(R-r_a)
  \;\le\;\sum_{a\in\{0,1\}^{n}}\mult_a P\;\le\;R\cdot2^{n-1}.
\]
Hence $\sum_a r_a\ge R\cdot2^{n-1}$ and
$\Ex_{a\in\{0,1\}^{n}}[r_a]\ge R/2$. In particular, there exists a binary point
$a$ with $\rank_{\F}N(a)\ge R/2$.

Now let $k\ge2$ and assume the conclusion holds for matrices in $k-1$ blocks.
Put $R=R(N)$. Apply the case $k=1$ with base field
$\F'=\F(\mathbf t_2,\dots,\mathbf t_k)$ to $N$, viewed as affine-linear in the
single block $\mathbf t_1$ with entries in $\F'[\mathbf t_1]$.
Its generic rank over $\F'(\mathbf t_1)=K$ is $R$, so there is a binary assignment
$\mathbf a_1$ with
\[
  \rank_{\F(\mathbf t_2,\dots,\mathbf t_k)}
  N(\mathbf a_1,\mathbf t_2,\dots,\mathbf t_k)\;\ge\;R/2.
\]
The degree bound and the constant matrix $C$ in the single-block argument are now
relative to $\F'$, which is legitimate because $\F'$ is a field.
The resulting matrix is multi-affine in the remaining $k-1$ blocks over $\F$.
By the induction hypothesis, it has a fully binary evaluation of rank at least
$2^{-(k-1)}\cdot R/2=2^{-k}R$. Thus $M(N)\ge2^{-k}R$, as required.
\end{proof}

\section*{AI methodology}

The author used GPT Astra in an extended dialogue to assist
with identifying and reviewing related work, exploring and refining proof strategies, and generating candidate proofs. This assistance materially influenced the related-work discussion in Section~1 and the development of the arguments presented in Section~\ref{sec:main-proof} and the supporting appendices. The author checked the mathematical arguments and the accuracy and attribution of the cited results, and substantially revised and reorganised the resulting material. The author takes full responsibility for the correctness, originality, and integrity of the submission.

\end{document}